    \documentclass[aps,prd,amsmath,twocolumn,groupedaddress,eqsecnum,showpacs,showkeys,nofootinbib]{revtex4}

 \def\fnote#1{\footnote}

 \renewcommand{\d}{{\rm d}} 
 \newcommand{\td}[2]{\frac{{\rm d} {#1}}{{\rm d} {#2}}}
 \newcommand{\tdil}[2]{{\rm d} {#1} / {\rm d} {#2}}
 \newcommand{\pd}[2]{\frac{\partial {#1}}{\partial {#2}}}
 \newcommand{\pdil}[2]{\partial {#1} / \partial {#2}}

 \usepackage[dvips]{graphicx}
 \usepackage{amsmath}
 \usepackage{amsfonts}
 \usepackage{amssymb}

\begin{document}

 \title{Formation of a galaxy with a central black hole in the
 Lema\^{\i}tre-Tolman model}

 \author{Andrzej Krasi\'{n}ski}
 \affiliation{N. Copernicus Astronomical Center,
 Polish Academy of Sciences, \\
 Bartycka 18, 00 716 Warszawa, Poland}
 \email{akr@camk.edu.pl}

 \author{Charles Hellaby}
 \affiliation{Department of Mathematics and Applied Mathematics, \\
 University of Cape Town, Rondebosch 7701, South Africa}
 \email{cwh@maths.uct.ac.za}

 \date {25/9/2003}

 \pacs{98.80.-k, 98.62.Ai, 98.62.Js}

 \keywords{Cosmology, Formation of galaxies, Black holes}

 \begin{abstract}
   We construct two models of the formation a galaxy with a central black
hole, starting from a small initial fluctuation at recombination. This is
an application of previously developed methods to find a
 Lema\^{\i}tre-Tolman model that evolves from a given initial density or
velocity profile to a given final density profile.  We show that the black
hole itself could be either a collapsed object, or a
 non-vacuum generalisation of a full
 Schwarzschild-Kruskal-Szekeres wormhole. Particular attention is paid to
the black hole's apparent and event horizons.
 \end{abstract}

 \maketitle

 \section{Aim and Motivation}

   It has become generally accepted that most large galaxies contain
central black holes (e.g. \cite{AnRvBlHoGalCen}--\cite{Bege2003}). This is due
to mounting evidence for very high luminosities within very small radii at the
centre of our galaxy and many others, as well as high orbital velocities of
stars very close to the centre, and is bolstered by observations of radio and
optical jets, as well as Seyfert galaxies and quasars at large redshifts.  Since
the `very small' radii accessible by current observations are still well outside
the horizons of the putative black holes, their exact nature, and even their
existence, is open to debate.  However, since the mean density inside the
horizon of a black hole is $\rho_s = 3 c^6 / 32 \pi G^3 M_s^2 = 1.845 \times
10^{17} / (M_s/M_\odot)^2$~g/cm$^3$, it is difficult to see how black hole
collapse can be avoided above $10^7~M_\odot$.

   The Lema\^{\i}tre-Tolman
 (L-T) model describes the behaviour of a spherically symmetric dust
distribution and has been a very fruitful source of models of inhomogenous
cosmology, smaller scale structure formation, and even black holes and
naked singularities.

   In paper I \cite{KrHe2002}, we considered the problem of finding a
spherically symmetric model that evolved from a given initial density
profile to a given final density profile.  We showed that this can always
be done with an
 L-T model, and we developed an alogorithm to find the arbitrary
functions of such an
 L-T model from the given profiles.  A numerical
example produced an Abell cluster from a density fluctuation at
recombination.

   In paper II \cite{KrHe2004a}, we generalised to finding
 L-T models that evolve from a
given velocity profile to a given density profile, the converse, and also
between two velocity profiles.  Several numerical examples, including the
evolution of a void, demostrated the usefulness of the method.

   We now utilise these methods to consider the formation of a galaxy with
a central black hole, a task for which the
 L-T model is particularly well
suited.  Although spiral galaxies are not exactly spherically symmetric,
both the core and the halo
 --- together containing more mass than the disk
 --- are quite close to it, so the
 L-T model is not a bad first
approximation.

The present state of the galaxy is defined by a mass distribution that consists
of two parts:

1. The part outside the apparent horizon at $t_2$ -- for which we use an
approximation to the observationally determined density profile of the M87
galaxy. This part extends inward to a sphere of mass $M_{BH}$, where $M_{BH}$ is
the observationally determined mass of the black hole in the M87 galaxy.

2. The part inside the apparent horizon at $t_2$. Since, for fundamental
reasons, no observational data at all exist for this region apart from the value
of $M_{BH}$, we were free to choose any geometry. We chose two examples:

2a. A simple subcase of the L--T model, discussed in sec. III F as an
illustrative example of properties of horizons. In this model, the black hole
does not exist initially and is formed in the course of evolution.

2b. A pre-existing wormhole, also chosen arbitrarily for simplicity of the
calculations.

\noindent The boundary between the ``inside'' and ``outside'' at times other
than $t_2$ goes along a comoving mass shell, so that at $t < t_2$ the apparent
horizon resides in the inside part.

For the initial state, at $t_1 =$ (the last scattering of the cosmic microwave
background (CMB) radiation), no usable observational data are available, either,
but hopefully only temporarily. The expected angular size on the CMB sky of a
perturbation that will develop into a single galaxy (0.004$^{\circ}$) is much
smaller than the current best resolution (0.2$^{\circ}$). Therefore we tried an
exactly homogeneous initial density and a homogeneous initial velocity. The
former turned out to lead to an unacceptable configuration at $t_2$: a
collapsing hyperbolic model with no Big Bang in the past. Consequently, we
settled on the homogeneous initial velocity, which then implied the amplitude
below $10^{-5}$ for the initial density perturbation.

These two states, at $t_1$ and $t_2$, uniquely define the L--T model that
evolves between them, as shown in Paper II. The 3-d surface graphs of density as
a function of mass and time show that the evolution proceeds without shell
crossings, and so the model is acceptable, at least qualitatively.

 \section{Basic properties of the Lema\^{\i}tre-Tolman model}

   The Lema\^{\i}tre-Tolman
 (L-T) model \cite{Lema1933, Tolm1934} is a spherically symmetric
nonstatic solution of the Einstein equations with a dust source.  See
\cite{Kras1997} for an extensive list of properties and other work on this
model.  Its metric is:
 \begin{equation}
   \d s^2 = \d t^2 - \frac {(R')^2}{1 + 2E(r)}\d r^2 -
   R^2(t,r)(\d\vartheta^2 + \sin^2\vartheta\d\varphi^2),
 \end{equation}
 where $E(r)$ is an arbitrary function of integration, $R' = \partial R/
\partial r$, and $R(t,r)$ obeys
 \begin{equation}\label{Rtsq}
   \dot{R}^2 = 2E + \frac{2M}{R} + \frac{\Lambda}{3} R^2,
 \end{equation}
 where $\dot{R} = \partial R / \partial t$ and $\Lambda$ is the
cosmological constant. Eq.~(\ref{Rtsq}) is a first integral of the
Einstein equations, and $M = M(r)$ is another arbitrary function  of
integration. The
 mass-density is:
 \begin{equation}   \label{rhoLT}
 \kappa \rho = \frac {2M'}{R^2R'}, \qquad
 \text{where}\ \kappa = \frac {8\pi G} {c^4}.
 \end{equation}
 In the following, we will assume $\Lambda = 0$. Then eq.~(\ref{Rtsq}) can
be solved explicitly, and the solutions are: when $E < 0$ (elliptic
evolution):
 \begin{subequations}
 \label{EllEv}
 \begin{eqnarray}
   \label{EllEvR}
   R(t,r) &= \frac{M}{(-2E)}(1 - \cos\eta),\\
   \label{EllEvt}
   \eta - \sin\eta &= \frac {(-2E)^{3/2}}{M} (t - t_B(r)),
 \end{eqnarray}
 \end{subequations}
 where $\eta$ is a parameter; when $E = 0$ (parabolic evolution):
 \begin{equation}   \label{ParEv}
   R(t,r) = \left[ \frac{9}{2} M (t - t_B(r))^2\right]^{1/3},
 \end{equation}
 and when $E > 0$ (hyperbolic evolution):
 \begin{subequations}
 \label{HypEv}
 \begin{eqnarray}
   R(t,r) &= \frac{M}{2E}(\cosh\eta - 1), \\
   \sinh\eta - \eta &= \frac {(2E)^{3/2}}{M} (t - t_B(r)),
 \end{eqnarray}
 \end{subequations}
 where $t_B(r)$ is one more arbitrary integration function (the bang
time). Note that all the formulae given so far are covariant under arbitrary
coordinate transformations $\tilde{r} = g(r)$, and so $r$ can be chosen at will.
This means one of the three functions $E(r)$, $M(r)$ and $t_B(r)$ can be fixed
at our convenience by the appropriate choice of $g$. We can define a scale
radius and a scale time for each worldline with
 \begin{eqnarray}
   \label{ScaleR}
   P(r) & = & \frac{2 M}{|2 E|} \\
   \label{ScaleT}
   T(r) & = & \frac{2 \pi M}{|2 E|^{3/2}}
 \end{eqnarray}
 and it is evident from (\ref{EllEv}) that, for the elliptic case, these
are the maximum $R$ and the lifetime for each $r$ value.  The crunch time is
then
 \begin{equation}\label{t_C}
   t_C(r) = t_B(r) + T(r) .
 \end{equation}

Writing eq. (\ref{EllEvt}) at $\eta = 2\pi$, where $t = t_C$, i.e. at the Big
Crunch, and then dividing the two equations we obtain
 \begin{equation}\label{ratioeta}
   \eta - \sin\eta = 2\pi (t - t_B)/(t_C - t_B),
 \end{equation}
so larger $\eta$ means only that the dust particle has completed a larger
fraction of its lifetime between the Bang and the Crunch.

 The parametric solutions (\ref{EllEv}) and
(\ref{HypEv}) can also be written
 \begin{widetext}
 \begin{align}
   \label{t_of_R_Ell_exp}
      t & = t_B + \frac{M}{(-2 E)^{3/2}} \left\{
      \arccos \left( 1 + \frac{2 E R}{M} \right)
      - 2 \sqrt{ \frac{- E R}{M} \left( 1 + \frac{E R}{M} \right) }\; 
      \right\},~~~~~~~~ 0 \leq \eta \leq \pi , \\
   \label{t_of_R_Ell_col}
      t & = t_B + \frac{M}{(-2 E)^{3/2}} \left\{ \pi +
      \arccos \left( - 1 - \frac{2 E R}{M} \right)
      + 2 \sqrt{ \frac{- E R}{M} \left( 1 + \frac{E R}{M} \right) }\;
      \right\} ,~~ \pi \leq \eta \leq 2 \pi ,
 \end{align}
 \end{widetext}
 for the expanding and collapsing elliptic cases, and
 \begin{multline}\label{t_of_R_Hyp}
   t = t_B + \frac{M}{(2 E)^{3/2}} \Bigg\{ 
   \sqrt{ \frac{2ER}{M} \left( 2 + \frac{2ER}{M} \right) }\; \\
   - {\rm arcosh} \left( 1 + \frac{2ER}{M} \right)
   \Bigg\}
 \end{multline}
 for the hyperbolic case (expanding).

   Apart from extended parabolic regions, there are also parabolic
boundaries between elliptic and hyperbolic regions, where $E \to 0$, but
$E' \neq 0$.  The limiting forms of equations (\ref{EllEv}) and
(\ref{HypEv}) are found by requiring well behaved time evolution and
setting
 \begin{equation}
   \eta = \tilde{\eta}\sqrt{E}\;
 \end{equation}
 so that $\tilde{\eta}$ is finite if $(t - t_B)$ is.

   The Friedmann models are contained in the
 L-T class as the limit:
 \begin{equation}\label{FriedLim}
   t_B = \text{const}, \qquad |E|^{3/2}/M = \text{const},
 \end{equation}
 and one of the standard radial coordinates for the Friedmann model
results if the coordinates in (\ref{EllEv})
 -- (\ref{HypEv}) are additionally chosen so that:
 \begin{equation}
   M = M_0 r^3 ~~~~\rightarrow~~~~  E = E_0 r^2
 \end{equation}
 with $M_0$ and $E_0$ being constants.

In constructing our galaxy model, it will be convenient to use $M(r)$ as the
radial coordinate (i.e. $\tilde{r} = M(r)$) --- because in most sections we
shall not need to pass through any ``necks" or ``bellies". Thus, $M(r)$ will be
a strictly growing function in the whole region under consideration. In some of
the sections we shall consider a black hole with a ``neck" or ``wormhole", but
even there, because of spherical symmetry, we will consider only one side of the
wormhole, where $M(r)$ is also increasing.

  Then with $R = R(t,M)$:
 \begin{equation}   \label{rhoRM}
   \kappa \rho = \frac{2}{R^2 \pd{R}{M}} \equiv \frac{6}{\pd{(R^3)}{M}} .
 \end{equation}

In the present paper we will apply the L-T model to a problem related to that
considered in papers I \& II: Connecting, by an L-T evolution, an initial state
of the Universe, defined by a mass-density or velocity distribution, to a final
state defined by a density distribution that contains a black hole.

 \subsection{Origin conditions}

   An origin, or centre of spherical symmetry, occurs at $r = r_c$ if
$R(t,r_c) = 0$ for all $t$.  The conditions for a regular centre have been
derived in \cite{MusHel2001} from the requirements that, away from the bang
and crunch, and in the limit $r \rightarrow r_c$:
 \begin{itemize}
 \item   $\eta$ in (\ref{EllEv}) and (\ref{HypEv}) must be finite if $\left(t -
t_B\right)$ is finite,
 \item   the density (\ref{rhoLT}) and the Kretschmann scalar are
 non-divergent, and the density is not zero,
 \item   on a constant time slice $d\rho/dR = 0$.
 \end{itemize}
In the equations below, the symbol $O_d(M)$ will denote a function that
has the property $\lim_{M \to 0} \left(O_d(M)/M^d\right) = 0$.  The
resulting conditions for the neighbourhood of $r_c$ are
 \begin{subequations}
 \label{OrigCondits}
 \begin{align}
   \label{OrigCond_R}
   & R = \beta(t) M^{1/3} +O_{1/3}(M)
      \mbox{~~~~along constant~} t , \\
   \label{OrigCond_E}
   & E = \gamma M^{2/3} + O_{2/3}(M), \\
   \label{OrigCond_tB}
   & t_B = \tau O_c(M) ~,~~~~ c > 1/3 , \\
   \label{OrigCond_rho}
   & \kappa \rho = 6/\beta^3 + O_0(M) , \\
   \label{OrigCond_M}
   & M(r_c) = 0.
 \end{align}
 \end{subequations}
 We also need $\tau < 0$ to avoid shell crossings.

 \subsection{Shell crossings, maxima and minima}

 Shell crossings, where a constant $r$ shell collides with its neighbour,
are loci of $R' = 0$ that are not regular maxima or minima of $R$.  They create
undesireable singularities where the density diverges and changes sign.  The
conditions on the 3 arbitrary functions that ensure none be present anywhere in
an L-T model, as well as those for regular maxima and minima in spatial
sections, were given in \cite{HeLa1985}, and will be used below.

 \section{Apparent and event horizons in the L-T model}

   We will be modelling a galactic black hole, so it will be useful to
consider its horizons.  Apparent and event horizons of
 L-T models were studied in \cite{Hell1987}, in which
 L-T models that generalise the Schwarzschild-Kruskal-Szekeres topology to
 non-vacuum were demonstrated.  It was shown that, when there's matter
present, the light rays get even less far through the wormhole than in the
vacuum case.  The diversity of possible topologies was discussed.  We lay out
further details of the apparent horizon here.

 \subsection{Definitions and basic properties}

   Let us write the evolution equation (\ref{Rtsq}) with $\Lambda = 0$ as
 \begin{multline}
   \dot{R} = \ell \sqrt{\frac{2 M}{R} + 2E}\; ~, \\
   \mbox{where}~
   \begin{cases}
   & \ell = +1 ~~~~\mbox{in the expanding phase,} \\
   & \ell = -1 ~~~~\mbox{in the collapsing phase.}
   \end{cases}
 \end{multline}
 The radial light rays must be geodesics by symmetry:
 \begin{equation}
   0 = -dt^2 + \frac{(R')^2}{1 + 2E} \, dr^2 ,
 \end{equation}
 and this may be written as
 \begin{multline}\label{dtdr_n}
   t'_n = \left. \td{t}{r} \right|_n = \frac{j R'}{\sqrt{1 + 2E}\;} ~, \\
   \mbox{where}~
   \begin{cases}
   & j = +1 ~~~~\mbox{for outgoing rays,} \\
   & j = -1 ~~~~\mbox{for incoming rays,}
   \end{cases}
 \end{multline}
 whose solution we write as $t = t_n(r, t_{n0})$, or often just $t_n(r)$
or $t_n$.

 \subsubsection{Apparent horizons}

 Along a ray we have
 \begin{eqnarray}
   R_n & = & R(t_n, r), \\
   \label{dRn_dr}
      (R_n)' & = & \dot{R} \left. \td{t}{r} \right|_n + R' \nonumber \\
      & = & \left( \ell j
      \frac{\sqrt{\frac{2 M}{R} + 2E}\;}{\sqrt{1 + 2E}\;}
      + 1 \right) R'.
 \end{eqnarray}
 The apparent horizon (AH) is the hypersurface in spacetime where the rays
are momentarily at constant $R$:
 \begin{gather}
   (R_n)' = 0 ~~~~\Rightarrow ~~~~
      \sqrt{\frac{2 M}{R} + 2E}\; = - \ell j \sqrt{1 + 2E}\;
      ~~~~\Rightarrow \\
   \label{eqaphor}   \ell j = -1, \qquad \mbox{and} \qquad R = 2M.
 \end{gather}
 There are in fact two apparent horizons: \\
 The future AH:~~(AH$^+$), where
 \begin{subequations}
 \begin{align}
   & \begin{cases}
        j = +1 ~~~~ & \mbox{(outgoing rays)} \\
        \ell = -1 ~~~~ & \mbox{(in a collapsing phase),}
     \end{cases}
 \intertext{and the past AH:~~(AH$^-$), where}
   & \begin{cases}
        j = -1 ~~~~ & \mbox{(incoming rays)} \\
        \ell = +1 ~~~~ & \mbox{(in an expanding phase).}
     \end{cases}
 \end{align}
 \end{subequations}
 We find $dt/dr$ along the AH by differentiating (\ref{eqaphor}):
 \begin{equation}
   \dot{R} \, dt + R' \, dr = 2 M' \, dr,
 \end{equation}
 giving
 \begin{equation}
   t_{\text{AH}}' = \left. \td{t}{r} \right|_{\text{AH}}
   = \frac{2 M' - R'}{\dot{R}}
   = \frac{2 M' -R'}{\ell \sqrt{\frac{2 M}{R} + 2E}\;},
 \end{equation}
 and, since $R = 2M$ on the AH,
 \begin{equation}\label{dtdr_AH}
   \left. \td{t}{r} \right|_{\text{AH}}
   = \frac{\ell (2 M' - R')}{\sqrt{1 + 2E}\;} .
 \end{equation}
 In the vacuum case $\rho = 0$, which implies $M' = 0$, we have
$\tdil{t}{r}|_{\text{AH}} = \tdil{t}{r}|_n$ since $\ell j = -1$.  Note
that $M' = 0$ could be only local, so the AH would only be null in that
region. In the Schwarzschild metric, where $M' = 0$ everywhere, this is
consistent with $R = 2M$ being the locus of the event horizons; and in
this case they coincide with the apparent horizons.

   Recall that in the Schwarzschild spacetime the future \& past event
horizons, EH$^+$ \& EH$^-$, cross in the neck at the moment it is widest.
(Call this event {\rm O}.)  This holds for
 L-T models too.  For hyperbolic regions, with $E \geq 0$ along each dust
worldline, there is either only expansion or only collapse, i.e. only one
AH (either AH$^+$ or AH$^-$) can occur. The AHs can thus cross only in an 
elliptic $E < 0$ region. At the neck of a
 L-T wormhole, where $2E = -1$, $M$ is a minimum, and $t_B$ is maximum,
the moment of maximum expansion is
 \begin{equation}
   \dot{R}^2 = 0 = \frac{2 M}{R} - 1 ~~~~\rightarrow~~~~
   R_{max}(M_{\text{min}}) = 2M.
 \end{equation}
 At all other $E$ values in an elliptic region $-1 < 2E < 0$, we find
$\dot{R} = 0 ~~~\rightarrow~~~ R_{max} = 2M/(- 2E) > 2M$. Thus $R = 2M$
has two solutions --- one in the expanding phase \& one in the collapsing
phase. So the AH$^+$ \& AH$^-$ meet at the neck maximum (event {\rm O}).

   To establish whether AH is timelike, null or spacelike, we compare the
slope of the AH$^+$ with the outgoing light ray (or the AH$^-$ with the
incoming light ray), i.e. $\ell j = -1$:

 \begin{equation}\label{DefOfB}
   B = \left. \left. \td{t}{r} \right|_{AH}\right
   / \left. \td{t}{r} \right|_{n}
   = - \ell j \left( 1 - \frac{2 M'}{R'} \right)
   = \left( 1 - \frac{2 M'}{R'} \right),
 \end{equation}
 but below, we will actually calculate
 \begin{multline}\label{timelikeAH}
   \overline{B} = \left( \frac{R'}{M'} - 1 \right)_{\text{AH}^\pm}
   = \left(\pd{R}{M} - 1 \right)_{\text{AH}^\pm} \\
   = \frac{1 + B}{1 - B}
   = 1 - \ell \sqrt{1 + 2 E}\; \td{t_{\text{AH}}}{M}
 \end{multline}
 where we have used (\ref{dtdr_AH}) and written $R'/M' = \pdil{R}{M}$ and
$t_B'/M' = \tdil{t_B}{M}$, since $M' > 0$.  Now since the conditions for
no shell crossings \cite{HeLa1985} require $M' \geq 0$ where $R' > 0$ and
 vice-versa, we have
 \begin{widetext}
 \begin{equation}\label{AHtype}
   \begin{array}{ccll}
   B_{max} = 1, & \overline{B} = + \infty
      & \rightarrow~\mbox{AH$^+$ outgoing null}
      & \mbox{(when~} M' = 0~,~ \rho = 0 \mbox{)} ,\\
   1 > B > -1, & + \infty > \overline{B} > 0
      & \rightarrow~\mbox{AH$^+$ spacelike}
      & \mbox{(for\ most~} M' \mbox{)} , \\
   B = -1, & \overline{B} = 0
      & \rightarrow~\mbox{AH$^+$ incoming null}
      & \mbox{(for\ large~} M'/R' \mbox{)} , \\
   -1 > B > -\infty, & 0 > \overline{B} > -1
      & \rightarrow~\mbox{AH$^+$ incoming timelike}
      & \mbox{(for\ very large~} M'/R' \mbox{)} ,
   \end{array}
 \end{equation}
 \end{widetext}
 so an outgoing timelike AH$^+$ is not possible.  This means outgoing
light rays that reach the AH$^+$ always fall inside AH$^+$, except where
$M' = 0$, in which case they move along it.  The possibility that AH$^\pm$
is timelike ($\overline{B} < 0$) holds if%
 \footnote{
 Note that eq. (\ref{dtAHdM_timelike}) can be equivalently written as
follows:
 $$
    \ell t'_{AH} > \frac {M'} {\sqrt{1 + 2E}} = {\cal M}',
 $$
where ${\cal M}$ is the sum of all the rest masses within the $r =$ const
sphere, equal to $4 \pi \int_0^r \sqrt{- g(t,x)}{\rm d}x \equiv 4 \pi \int_0^r
\left[R^2(t,x)R'(t,x)/\sqrt{2E + 1}\right]{\rm d}x$, while $M = 4 \pi \int_0^r
R^2(t,x)R'(t,x){\rm d}x$ is the active gravitational mass (see \cite{Bond1947}
for a detailed discussion). }
 \begin{equation}\label{dtAHdM_timelike}
   \ell \td{t_{\text{AH}}}{M} > \frac{1}{\sqrt{1 + 2 E}\;} .
 \end{equation}
 This means that the smaller $E$ is, the steeper the locus of the AH must
be to make it timelike.

The argument is similar for light rays at AH$^-$, except that `incoming' should
swopped with `outgoing'.  If ingoing light rays reach the AH$^-$, they pass out
of it or run along it.

   Since this is true for every point on AH$^+$ \& AH$^-$, radial light
rays in dense wormholes are more trapped and go even less far than in
vacuum.  In particular, if%
 \footnote{
 Since a regular minimum or maximum $R' = 0$ requires all of $M'$, $E'$
and $t_B'$ to be locally zero, $\rho > 0 \Rightarrow M''/R'' > 0$.
 }
 $\rho > 0$ where $2E = -1$, light rays starting at {\rm O} fall inside
AH$^+$ to the future and inside AH$^-$ to the past.

 \subsubsection{Event horizons}

   The event horizon is the very last ray to reach future null infinity
(EH$^+$), or the very first one to come in from past null infinity (EH$^-
$).  If we have vacuum ($M' = 0$) everywhere, then light rays travel along
$R = 2M$, and the EHs coincide with the AHs.  If there is matter, $M' >
0$, on any worldline, then the incoming light rays emerge from AH$^-$ and
outgoing light rays fall into AH$^+$ at that $r$ value, and so the EHs
split off from the AHs (see \cite{Hell1987}).

 \subsection{Locating the apparent horizons in elliptic regions}

 \subsubsection{AH$^-$ during expansion}

   We shall first consider the expansion phase of an elliptic model, where
$0 \leq \eta \leq \pi$ and $E < 0$, so we have $\ell = +1$ and only AH$^-$
is present.  Since $R = 2M$ on an AH, we have from (\ref{EllEvR}):
 \begin{equation}\label{etaonAH}
   \cos\eta_{\text{AH}} = 1 + 4E ,
 \end{equation}
 and thus, along a given worldline, the proper time of passing through the
AH, counted from the Bang time $t_B$, can be calculated from
(\ref{t_of_R_Ell_exp}) with $R = 2M$ to be
 \begin{equation}\label{timeleaveAH}
   t_{\text{AH}^-} - t_B = M
   \frac {\arccos\left(1 + 4E \right) - 2 \sqrt{- 2E\left(1 + 2E\right)}}
   {\left(- 2E \right)^{3/2}}.
 \end{equation}
The function $F = t_{\text{AH}^-} - t_B$ of the argument $f = 2E$, defined in
(\ref{timeleaveAH}) has the following properties
 \begin{equation}\label{proptimeleaveAH}
   \begin{split}
   F(-1) &= M\pi, \qquad F(0) = 4M/3, \\
     \frac {{\rm d} F} {{\rm d}f} &< 0 \qquad {\rm for\ \ } -1 < f < 0,
   \end{split}
 \end{equation}
 i.e. it is decreasing.  These properties mean that the AH does not touch
the Big Bang anywhere except at a centre, $M = 0$, even if $E = 0$.

 Along all worldlines with $E > -1/2$, even parabolic worldlines $E = 0$
\& $M \neq 0$, the dust particles emerge from AH$^-$ a finite time after the Big
Bang ($\eta = 0$) and a finite time before maximum expansion ($\eta = \pi$).  In
order to find the slope of AH$^-$, we differentiate (\ref{timeleaveAH}) with
respect to $M$ to obtain
 \begin{equation}\label{dertonAH}
 \begin{split}
   \td{t_{\text{AH}^-}}{M} = & \left[\frac{1}{(-2E)^{3/2}}
      + \frac{3M}{(-2E)^{5/2}} \td{E}{M} \right]
      \arccos\left(1 + 4E\right) \\
   & + \frac{1}{E} \sqrt{1 + 2E} - \frac {M(3 + 2E)} {2E^2\sqrt{1 + 2E}}
      \td{E}{M}  + \td{t_B}{M} .
 \end{split}
 \end{equation}
 In general, this is very difficult to analyse, but for special cases it
will be possible.

So although (\ref{etaonAH}) shows wordlines with larger $E$ exit AH$^-$ at a
later stage of evolution, this may not correspond to a later time $t$, or even
to a longer time $(t - t_B)$ since the the bang.  It is not at all necessary
that $E$ is a monotonically decreasing function of $r$ in an elliptic region; in
general it can increase and decrease again any number of times.

 \subsubsection{AH$^-$ near an origin}

   On the AH$^-$ in the neighbourhood of a regular center, we obtain from
(\ref{etaonAH}) and (\ref{OrigCond_E}) to lowest order:
 \begin{equation}\label{etacloseto0}
   1 - \cos\eta_{\text{AH}} = -4 \gamma M^{2/3} + O_{2/3}(M).
 \end{equation}
 and consequently, $\eta_{\text{AH}^-} \to 0$ as $r \to r_c$, i.e. AH$^-$
touches the Big Bang set at the center.  Since $E(r_c) = 0$ and $E < 0$ in
the neighbourhood of the center, it follows that $E'(r_c) \leq 0$, and,
via (\ref{etaonAH}), that the dust particles with larger $r$ (smaller $E$)
exit AH$^-$ with larger values of $\eta$.

To show the behaviour of AH$^-$ near $r_c$ is not unique, we take the following
example for $E(M)$
 \begin{multline}\label{EnearOrig}
   E = M^{2/3} (\gamma + \gamma_2 M^d) ~, \\
   ~~~~ d > 2/3
   ~,~~~~ \gamma \neq 0
   ~,~~~~ \gamma_2 \neq 0 .
 \end{multline}
 Putting (\ref{EnearOrig}) and (\ref{OrigCond_tB}) in (\ref{dertonAH}) we
find%
 \footnote{
 Since $M^{1/3}$ is a natural measure of proper radius near an origin, the
slope of AH$-$ is zero:
 $$
   \td{t_{\text{AH}^-}}{M^{1/3}} \approx
   3 c \tau M^{c - 1/3}
   + 4 M^{2/3} + ... \approx 0 .
 $$
 Away from the centre, the sign of $\tdil{t_{\text{AH}^-}}{M^{1/3}}$ depends
on whether or not $c > 1$, as $\tau < 0$ for no shell crossings \cite{HeLa1985}.
 }%
 , neglecting powers of $M$ that are necessarily positive,
 \begin{equation}\label{CentGradtAH-}
\td{t_{\text{AH}^-}}{M} \approx c \tau M^{c - 1} + \frac{4}{3},
 \end{equation}
 and from (\ref{timelikeAH}) with $\ell = +1$ and
(\ref{CentGradtAH-}) we see that,
 \begin{equation}\label{B-bar-origin}
   \overline{B} \approx - c \tau M^{c - 1} - \frac{1}{3} + ... .
 \end{equation}
 The behaviour of the apparent horizon in the vicinity of the center is not
unique, each of the cases listed in (\ref{AHtype}) can occur, depending on
whether $c > 1$. This nonuniqueness of behaviour is connected with the
shell-focussing singularities that appear in some L-T models.  Various studies
\cite{EaSm1979}-\cite{Josh1993} have shown that outgoing light rays may emerge
from the central point of the Big Crunch and even reach infinity. Moreover, even
though, in those cases, this central point (where the Big Crunch first forms)
appears to be a single point in comoving coordinates, it is in fact a finite
segment of a null line in the Penrose diagram, see \cite{EaSm1979}.  At the Big
Bang, we have the reverse
 --- incoming light rays may reach the central point of the Bang
singularity.  Any radial light ray emitted from the center of symmetry
that falls into the Big Crunch must first increase its $R$ value, and then
decrease, i.e. must cross the AH+ in between. In consequence, the AH+
cannot touch the center of BC earlier than the null singularity does.
Similarly, the AH$^-$ cannot touch the center of BB later than the null
singularity does.

 \subsubsection{AH$^-$ in the parabolic limit}
 \label{AHParaLim}

   A shell of parabolic worldlines occurs at the boundary between elliptic
and hyperbolic regions, where $E \to 0$, but $E' \neq 0$ and%
 \footnote{
 Elliptic and parabolic regions are not possible for $M = 0$
 }
 $M > 0$.  From (\ref{etaonAH}), (\ref{timeleaveAH}), (\ref{ScaleT}),
(\ref{dertonAH}) and (\ref{timelikeAH}) with $\ell = +1$, we see that
 \begin{subequations}
 \begin{align}
   \eta & \xrightarrow[E \to 0]{} 0 \\
   \label{Para_Lim_tAH-}
   t_{\text{AH}^-} - t_B & \xrightarrow[E \to 0]{} \frac{4 M}{3} \\
   T & \xrightarrow[E \to 0]{} \infty \\
   \label{Para_Lim_dtAH-dM}
   \td{t_{\text{AH}^-}}{M} & \xrightarrow[E \to 0]{}
      \frac{4}{3} - \frac{4 M}{5} \td{E}{M} + \td{t_B}{M} \\
   \label{Para_Lim_BbarAH-}
   \overline{B}_{\text{AH}^-} & \xrightarrow[E \to 0]{}
      - \frac{1}{3} + \frac{4 M}{5} \td{E}{M} - \td{t_B}{M}
 \end{align}
 \end{subequations}
so AH$^-$ never touches the bang here, despite $\eta$ being zero. The divergence
of the worldline lifetime $T$ suggests that either the bang time or the crunch
time recedes to infinity, as would be expected in a hyperbolic region.  There is
in fact a third possibility, as there is no reason why both times should not
diverge.  Indeed, an asymptotically flat model is achieved by letting $E \to 0$
as $r \to \infty$ \cite{Hell1987},
with both bang and crunch times diverging%
 \footnote{Another example of this kind will be discussed in sec.
\ref{IllustrationModel}.}%
 .  We also see the slope and the causal nature of AH$^-$ are uncertain
here.

 \subsubsection{AH$^+$ during collapse}

 This can be obtained form the above by replacing $(t - t_B)$ with $(t_C -
t)$, $\eta$ with $(2 \pi - \eta)$, flipping the signs of $\ell$ and $j$,
and swopping ``incoming" with ``outgoing".  However, keeping $t_B$ as our
arbitrary function, we have $\pi \leq \eta \leq 2 \pi$.  Eq.
(\ref{etaonAH}) still applies, but instead of (\ref{timeleaveAH}) and
(\ref{dertonAH}), we now obtain
 \begin{gather}\label{timeleaveAHcoll}
   \begin{split}
      & (t - t_B)_{\text{AH}^+} = \\
      & M \frac {\pi + \arccos (-1 - 4E)
         + 2 \sqrt{- 2E (1 + 2E)}} {(- 2E)^{3/2}} ,
   \end{split} \\
   \begin{split}
      & \td{t_{\text{AH}^+}}{M} = \\
      & \left[\frac{1}{(-2E)^{3/2}}
         + \frac{3M}{(-2E)^{5/2}} \td{E}{M} \right]
         [\pi + \arccos\left(-1 - 4E\right)] \\
      & - \frac{1}{E} \sqrt{1 + 2E}
         + \frac {M(3 + 2E)} {2E^2\sqrt{1 + 2E}} \td{E}{M}
         + \td{t_B}{M} .
   \end{split}
 \end{gather}
 and (\ref{timelikeAH}) applies with $\ell = -1$.  The special cases all
follow in the same way.  Near an origin, we again find nonunique
behaviour, one of the possibilities this time being light rays escaping
the crunch at $r_c$ before AH$^+$ forms.  At a regular extremum where $E =
-1/2$, $\eta_{\text{AH}^+} \to \pi$ and AH$^+$ crosses AH$^-$.  In the
parabolic limit, $\eta_{\text{AH}^+} \to 0$, $t_C - t_{\text{AH}^+} \to
4M/3$, and
 \begin{subequations}
 \begin{align}
   \td{t_{\text{AH}^+}}{M} & \xrightarrow[E \to 0]{}
      - \frac{4}{3} + \frac{4 M}{5} \td{E}{M} + \td{t_C}{M} , \\
   \overline{B}_{\text{AH}^+} & \xrightarrow[E \to 0]{}
      - \frac{1}{3} + \frac{4 M}{5} \td{E}{M} + \td{t_C}{M} ,
 \end{align}
 \end{subequations}
where $t_C$ is given by (\ref{t_C}) \& (\ref{ScaleT}).

 \subsection{Apparent horizons in parabolic regions}

   The corresponding results in an expanding $E = 0$, $E' = 0$
 L-T model follow from section \ref{AHParaLim} (or directly from
(\ref{ParEv}) with $R = 2M$).  In particular (\ref{Para_Lim_dtAH-dM}) and
(\ref{Para_Lim_BbarAH-}) become
 \begin{subequations}\label{aphorpara}
 \begin{align}
   \td{t_{\text{AH}^-}}{M} & = \frac{4}{3} + \td{t_B}{M} \\
   \overline{B}_{\text{AH}^-} & = - \frac{1}{3} - \td{t_B}{M}
 \end{align}
 \end{subequations}
 Despite the no shell crossing condition $\tdil{t_B}{M} < 0$, AH$^-$ may
still exhibit all possible behaviours of (\ref{AHtype}) with ``outgoing''
and ``incoming'' interchanged. In a collapsing parabolic model, time
reversed results apply.

 \subsection{Locating apparent horizons in hyperbolic regions}

   Using the same methods as for expanding elliptic regions, we find the
behaviour of the AH in expanding hyperbolic regions is qualitatively the
same.  There is of course only one AH, no maximum expansion, and loci
where $E = -1/2$ are not possible.  But the results for origins and for
the parabolic limit both carry over.  Collapsing hyperbolic regions are
essentially like collapsing elliptic regions.

 \subsection{Locating the event horizon}
 \label{EHLoc}

   In general, locating the EH involves integrating (\ref{dtdr_n}), which
cannot be done analytically because $R'$ is not a simple expression.  In
addition, we must find the last outgoing ray to escape the crunch, or the
first incoming ray to avoid the bang.  For a particular model, EH$^+$ is
located numerically using a compactified coordinate representation of the
spacetime in section \ref{EHSec}.

 \subsection{An illustration --- a simple recollapsing model}
 \label{IllustrationModel}

   We shall illustrate several properties of the L-T model and of its
apparent horizons on a simple example. In the present subsection, we will
use this model with unrealistic parameter values, chosen in such a way
that all the figures are easily readable. Later, in sec. \ref{gala+bh}, we
will use the same model for modelling a galaxy with a black hole at the
center, with parameters chosen to fit observational data.

 \subsubsection{Definition}

   We take an $E < 0$
 L-T model with a regular centre, whose Big Bang function $t_B(M)$ is
 \begin{equation}\label{exbafu}
   t_B(M) = - b M^2 + t_{B0},
 \end{equation}
 and whose Big Crunch function is
 \begin{equation}\label{excrfu}
   t_C(M) = a M^3 + T_0 + t_{B0} ,
 \end{equation}
 where the parameter $T_0$ is the lifetime of the central worldline where
$M = 0$. The numerical values of the parameters used in the figures will be $a =
2\cdot 10^4$, $b = 200$, $t_{B0} = 5$, $T_0 = 0.05$.  Their values were chosen
so as to make the figures readable and illustrative, at this point they are
unrelated to any astrophysical quantities.  Since $t = t_C$ at $\eta = 2\pi$, we
find from eq. (\ref{EllEv}):
 \begin{align}\label{exener}
   E(M) & = - \frac 1 2 \left(\frac {2\pi M} {t_C - t_B}\right)^{2/3} 
      \nonumber\\
   & = - \left(\frac {\pi^2} 2\right)^{1/3}\
      \frac {M^{2/3}} {\left(aM^3 + bM^2 + T_0\right)^{2/3}}\ .
 \end{align}
 Note that as $M \to \infty$, we have $t_B \to - \infty$, $t_C \to +
\infty$ and $E \to 0$. Hence, the space contains infinite mass and has
infinite volume. Unlike in the Friedmann models, positive space curvature
does not imply finite volume; this has been known since long ago
\cite{Bonn1985, HeLa1985}.

 \subsubsection{Description}

\begin{figure}[b]
 \includegraphics[scale = 0.45]{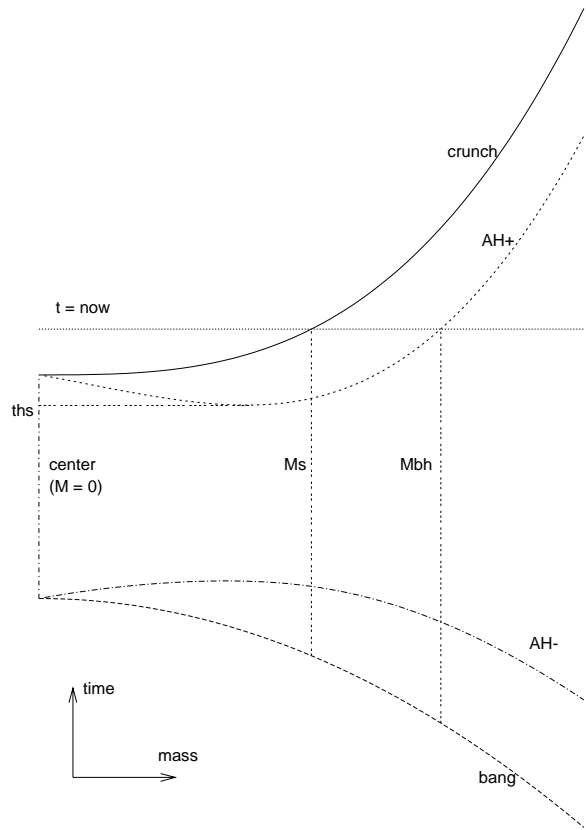}
 \caption{
 \label{bhspace}
 Evolution leading to a black hole in the $E < 0$ L-T model of eqs.
(\ref{exbafu}) -- (\ref{excrfu}). The final state is defined at the
instant $t = t_2 =$ now. Worldlines of dust particles are vertical
straight lines, each has a constant mass-coordinate. Intersections of the
line $t = t_2$ with the lines representing the Big Crunch and the future
apparent horizon determine the masses $M_S$ and $M_{BH}$, respectively.}
 \end{figure}

   The main features of this model are shown in Fig. \ref{bhspace}.  Note
that the AH$^+$ first appears not at the center, but at a finite distance
from the center, where the function $t_{AH^+}(M)$ has its minimum, and at
a time $t_{hs} < t_C(0)$.

 At all times after the crunch first forms, $t > t_C(0)$, the mass $M_S$
already swallowed up by the singularity is necessarily smaller than the mass
$M_{BH}$ that disappeared into the AH$^+$.  So for an object in which the black
hole already exists, the time $t = $ now must be taken after $t_{hs}$ and in
Fig. \ref{bhspace} it is also greater than $t_C(0)$. The mass $M_S$ cannot even
be estimated by astronomical methods.

   The situation is similar, but reversed in time, for the Big Bang
singularity and the AH$^-$.

\begin{figure}[b]
 \includegraphics[scale = 0.55]{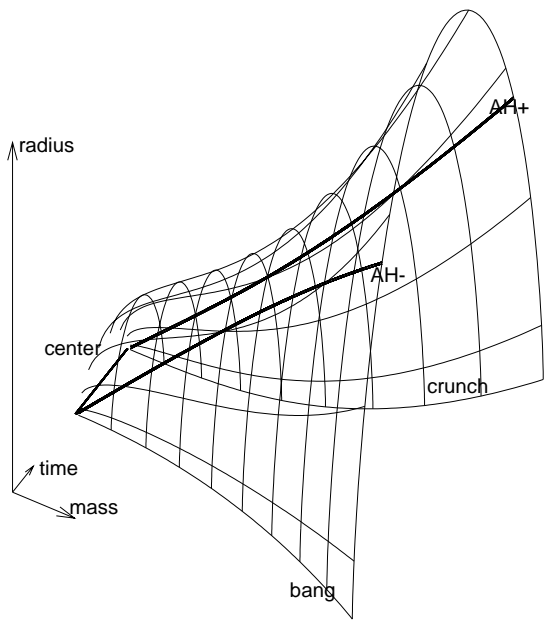}
 \caption{
 \label{bhspace4}
 3-d graph of the black hole formation process from Fig. \ref{bhspace}:
the areal radius as a function of $M$ and $t$. Each shell of constant mass
evolves in a plane given by $M=$~const. It starts at $R = 0$, then
gets out of the past apparent horizon AH$^-$, then reaches maximum $R$,
then falls into the future apparent horizon AH$^+$, and finally hits the
Big Crunch. Note that the surface intersects the $R = 0$ plane
perpendicularly all along the $R = 0$ contour. The apparent horizons are
intersections of the $R(M, t)$ surface with the plane $R = 2M$. (Hence, it
is easy to figure out how they would look in the $k > 0$ Friedmann model,
where $t_B = $ const, $t_C = $ const and all the $\left.R(t)\right|_{M =
\text{const.}}$ curves are indentical. }
 \end{figure}

\begin{figure*}[!]
 \includegraphics[scale = 0.95]{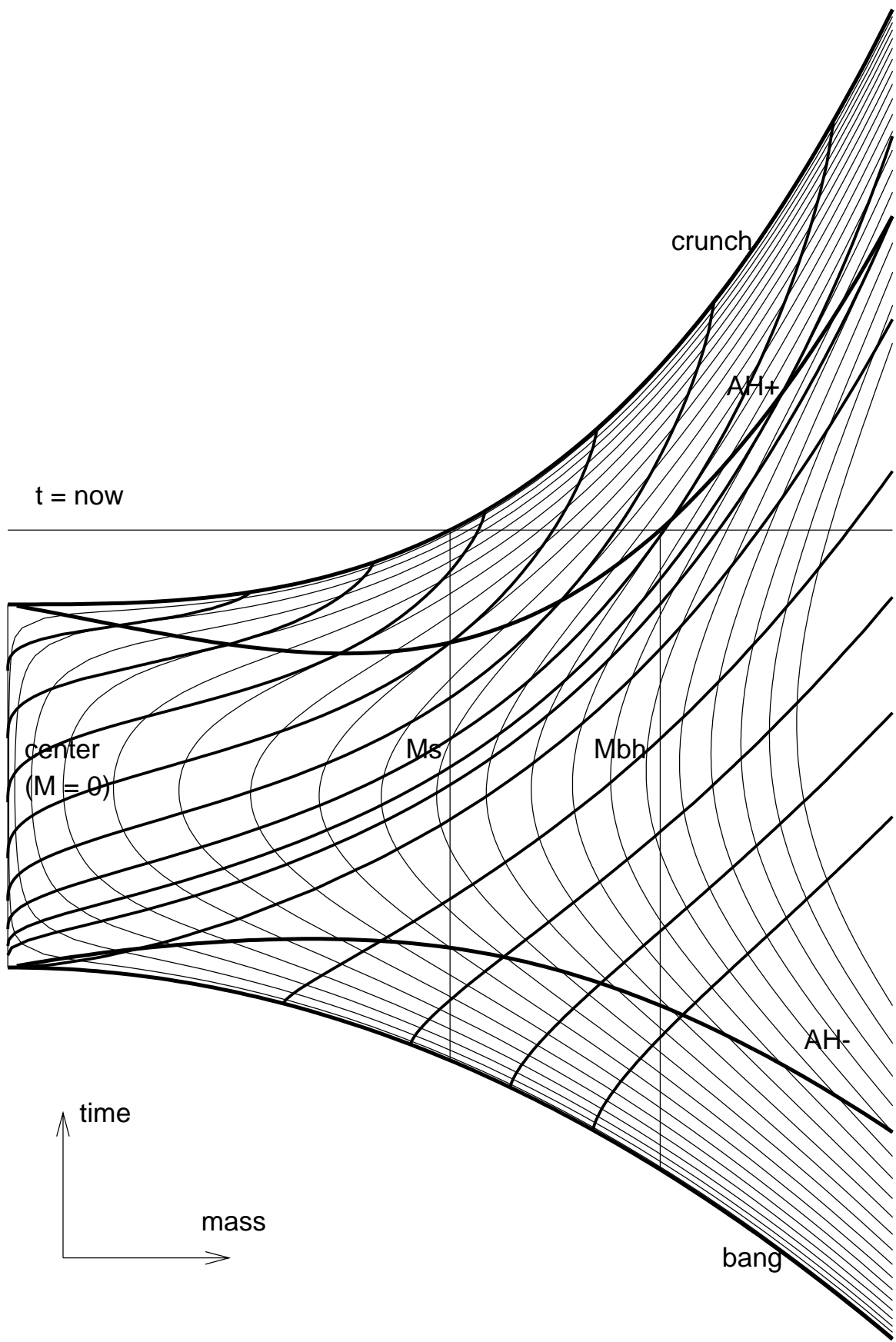}
 \caption{
 \label{allgeomap}
 Contours of constant $R$-value (thinner lines) and outgoing radial null
geodesics inscribed into the spacetime diagram of Fig. \ref{bhspace}. The
$R$-values on consecutive contours differ by always the same amount. }
 \end{figure*}

   The evolution of our model can be visualised more graphically in a 3-d
diagram (Fig. \ref{bhspace4}), which shows the value of the areal radius
$R$ at each time $t$ and each value of $M$.
 %
 %
 Fig. \ref{allgeomap} shows the ``topographic map'' of the surface from
Fig. \ref{bhspace4}. It contains contours%
 \footnote{
 Fig. \ref{allgeomap} can be used to explain some properties of shell
crossings. Since the worldlines of the dust source are constant $M$ lines
(i.e. vertical straight lines), they can never intersect in a
 $t$-$M$ diagram, even at shell crossings. A shell crossing would show in
this picture as a point where a
 constant-$R$-contour has a horizontal tangent. (No such points exist in
this case, because the functions of the
 L-T model had been chosen appropriately.) The figure graphically explains
why, for avoiding shell crossings away from a neck or belly, it is
necessary that the Big Bang is a decreasing function of $M$, and the Big
Crunch is an increasing function of $M$. The center of symmetry, the Big
Crunch and the Big Bang together form the $R = 0$ contour. Contours of
small constant $R$ must have a similar shape. Hence, if either of the two
conditions were not fulfilled, either the upper branch or the lower branch
of some contours would be a
 non-monotonic function, whose derivative by $M$ would change sign
somewhere. At the changeover points, the tangents to the contours would be
horizontal, and these would be the shell crossings.
 }
 of constant $R$ (the thinner curves) inscribed into Fig. \ref{bhspace}.
(The other curves are outgoing radial null geodesics, and we shall discuss them
further below.)

   Rembering that in the $t$--$M$ diagram, the slopes of the
incoming/outgoing radial light rays are each other's mirror images about
vertical lines, it is evident from Fig. \ref{allgeomap} that AH$^\pm$ are
spacelike everywhere except possibly in a neighbourhood of the central
line $M = 0$ and at future null infinity.  From (\ref{exener}) and
(\ref{exbafu}) we have $e = 8/3$ and $c = 2$ in (\ref{B-bar-origin}), so
we find $\overline{B}_{\text{AH}^-} \rightarrow 1$ at the origin.
Similarly by (\ref{excrfu}) we have $c = 3$, and
$\overline{B}_{\text{AH}^+ } \rightarrow 1$. Therefore AH$^\pm$ are both
spacelike at the origin too.
 %
 %

Fig. \ref{allgeomap} also shows several outgoing radial null geodesics. Each
geodesic has a vertical tangent at the center. This is a consequence of using
$M$ as the radial coordinate. Since $\tdil{t}{M} = \pdil{R}{M}/\sqrt{1 + 2E}$ on
each geodesic and $R \propto M^{2/3}$ close to the center, so $\tdil{t}{M}
\propto M^{-1/3}$ and $\tdil{t}{M} \to \infty$ as $M \to 0$. Each geodesic
proceeds to higher values of $R$ before it meets the apparent horizon AH$^+$. At
AH$^+$, it is tangent to an $R = $ const contour, then proceeds toward smaller
$R$ values. The future event horizon consists of those radial null geodesics
that approach the AH$^+$ asymptotically. In the figure, it lies between the
geodesics no 5 and 6, counted from the lower right corner of the figure; we
shall discuss its location in more detail in sec. \ref{EHSec}.

Geodesic no 5 from the lower right emanates from the center $M = 0$, where the Big
Bang function has a local maximum. The tangent to the geodesic is horizontal
there. This means that the observer receiving it sees the light infinitely
redshifted, as in the Friedmann models. Geodesics to the right of this one all
begin with a vertical tangent, which implies an infinite blueshift. These
observations about redshift and blueshift were first made by Szekeres
\cite{Szek1980}. Likewise, the geodesics meet the Big Crunch with their tangents
being vertical.

   By the time the crunch forms at $t = t_C(0)$, the future apparent
horizon already exists (see Figs. \ref{bhspace} and \ref{allgeomap}).  The
shells of progressively greater values of $M$ first go through the AH, and then
hit the singularity at $t = t_C(M)$. We assume that at the time $t = t_2$, the
singularity has already accumulated the mass $M_S$, while the mass hidden inside
the apparent horizon at the same time is $M_{BH} > M_S$. Both of them grow with
time, but at fixed $t_2$ they are constants. From the definitions of $M_S$ and
$M_{BH}$ it follows that
 \begin{equation}\label{msmbhat2}
   t_2 = t_C(M_S) = t_{AH+}(M_{BH}).
 \end{equation}

 \subsubsection{Location of the event horizon.}\label{EHSec}

   Now we shall discuss the location of the event horizon in the spacetime
model considered in sec. \ref{IllustrationModel}. It will follow that, even
though the model has a rather simple geometry, this is quite a complicated task
that requires complete knowledge of the whole spacetime, including the null
infinity. Hence, in a real Universe, where our knowledge is limited to a
relatively small neighbourhood of our past light cone and our past worldline,
and the knowledge is mostly incomplete and imprecise, the event horizon simply
cannot be located by astronomical observations.

\begin{figure}[b]
 \includegraphics[scale = 0.45]{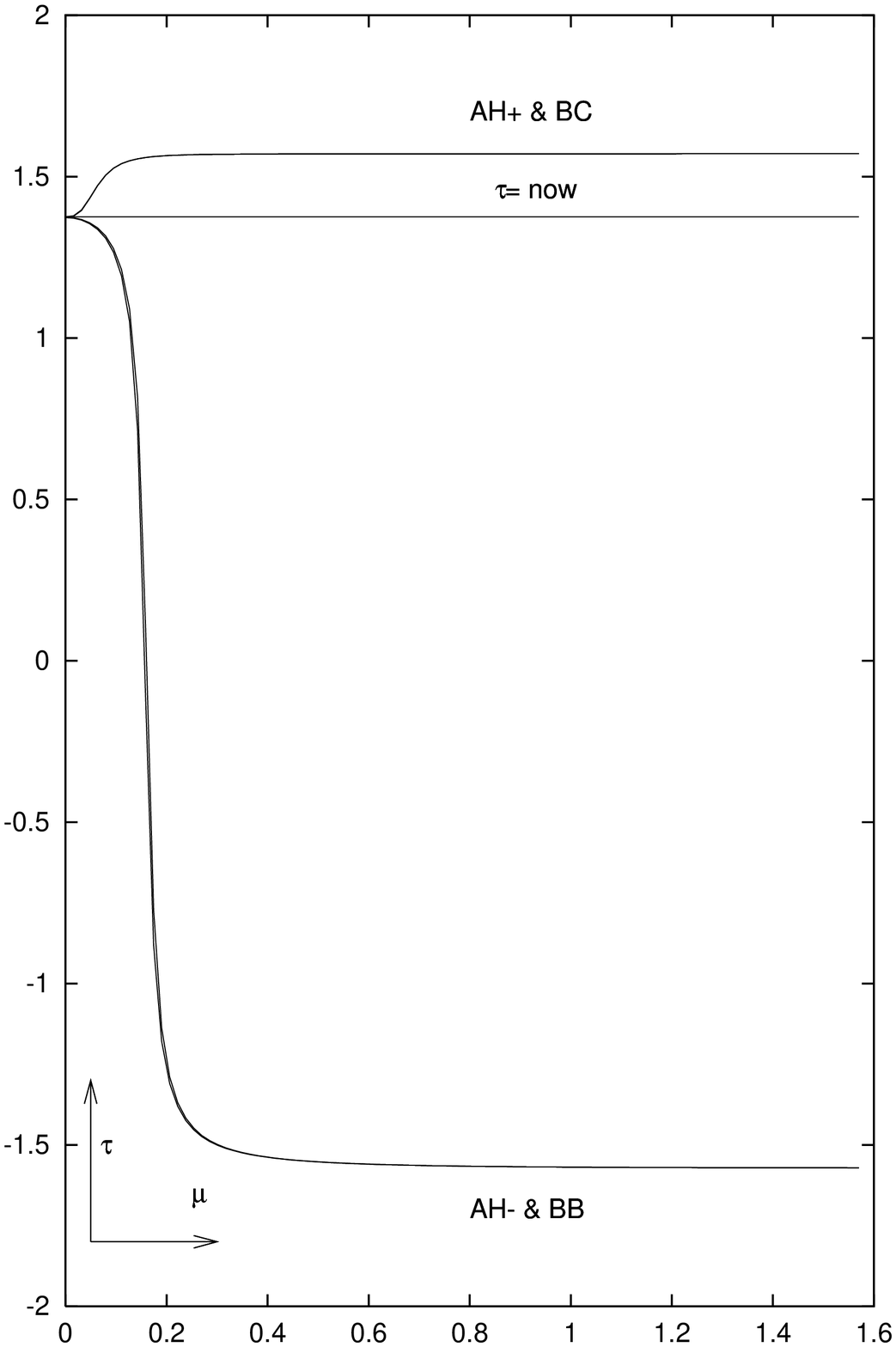}
 ${}$ \\[-90mm]
 \hspace*{18mm}
 \includegraphics[scale = 0.27]{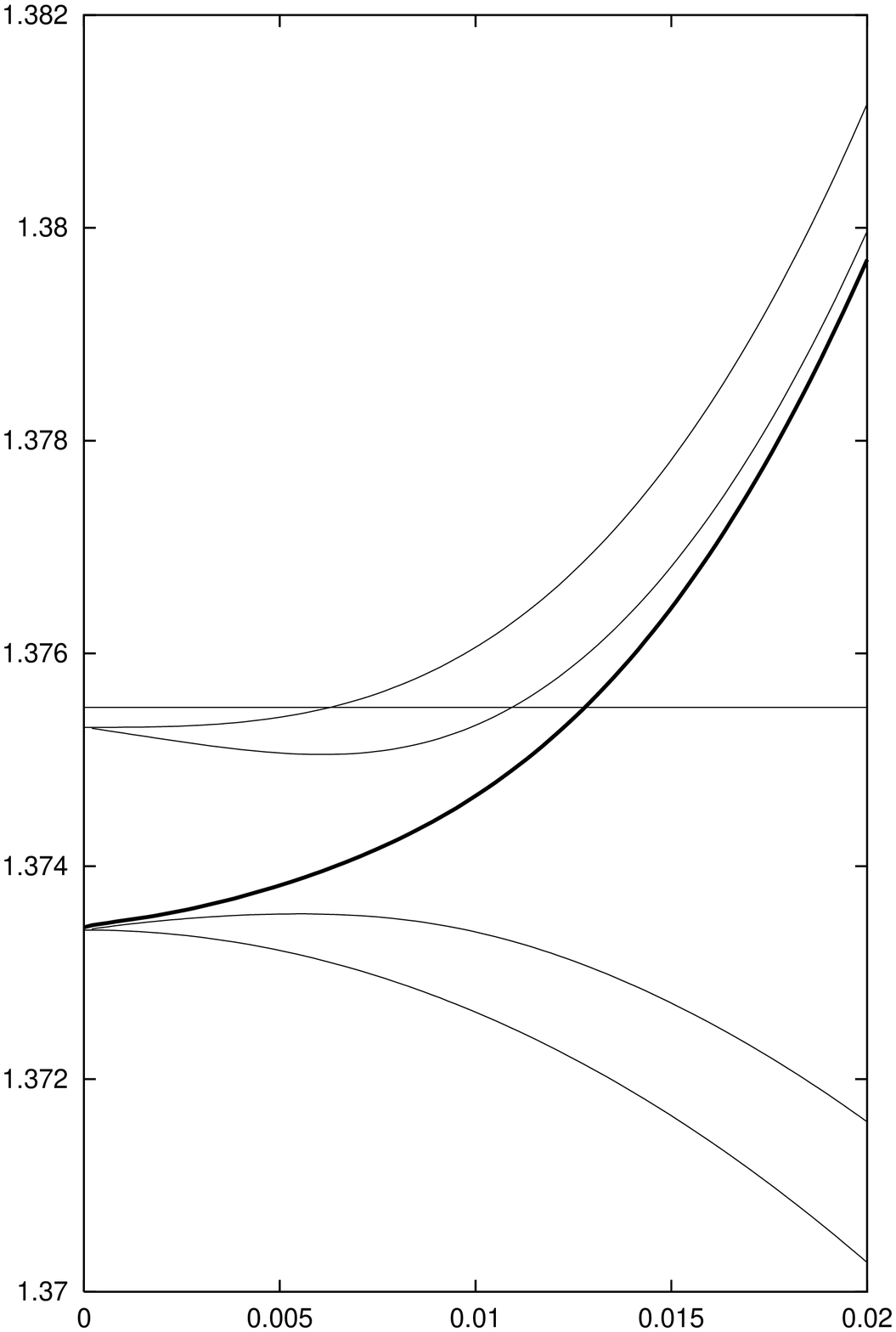}
 ${}$ \\[20mm]
 \caption{
 \label{compbhglob}
 The spacetime diagram of Fig. \ref{bhspace} compactified according to eq.
(\ref{compcoor}). The whole region shown in Fig. \ref{bhspace} is squeezed into
the point where the three lines meet at the $\tau$-axis. The worldlines of the
dust source would still be vertical straight lines here. The upper curve is the
future apparent horizon and the Big Crunch singularity, they seem to coincide at
the scale of this picture. The lower curve is the past apparent horizon and the
Big Bang singularity, again coinciding only spuriously. The horizontal straight
line is the $\tau = $ now time. Inset: a closeup view of the image (in the
coordinates $(\mu, \tau)$) of the region shown in Fig. \ref{bhspace}. The
thicker line is the event horizon. It does not really hit the central point of
the Big Bang; the apparent coincidence is just an artefact of the scale. More
explanation in the text.}
 \end{figure}

   The future event horizon is formed by those null geodesics that fall
into the future apparent horizon ``as late as possible'', i.e. approach it
asymptotically. Hence, in order to locate the event horizon, we must issue
null geodesics backward in time from the ``future end point'' of the AH+.
This cannot be done in the $(M, t)$ coordinates used so far because the
spacetime and the AH+ are infinite. Hence, we must first compactify the
spacetime. The most convenient compactification for considering null
geodesics at a null infinity is, theoretically, a Penrose transform
because it spreads the null infinities into finite sets. However, in order
to find a Penrose transformation, one must first choose null coordinates,
and in the
 L-T model this has so far proven to be an impossible task, see
Ref. \cite{StEN1992,Hell1996}. Hence, we will use a less convenient
compactification that will squeeze the null infinities into single points
in the 2-dimensional (time - radius) spacetime diagram. It is provided by
the transformation \begin{equation}\label{compcoor} M = \tan(\mu), \qquad
t = \tan(\tau). \end{equation} In these coordinates, the ${\mathbb R}^1
\times {\mathbb R}^1_+$ space of Figs. \ref{bhspace} -- \ref{allgeomap}
becomes the finite $[- \pi/2,\pi/2] \times [0, \pi/2]$ rectangle, see Fig.
\ref{compbhglob}. The upper curve in the figure is the Big Crunch
singularity; the future apparent horizon runs so close to it that it seems
to coincide with it%
 \footnote{
 As can be verified from eq. (\ref{timeleaveAHcoll}), the time-difference
between the crunch and the AH+ goes to infinity when $M \to \infty$.
However, the ratio of this time-difference to the crunchtime goes to zero,
which explains why the two curves in Fig. \ref{compbhglob} meet at the
image of the infinity. The same is true for the Big Bang and the AH$^-$.
 }%
 .  The horizontal line is the $\tau =$ now line. The lower curve is the Big
Bang and the past apparent horizon, again running one on top of the other.
The point on the $\tau$-axis where the three lines meet is the image of
the $M = 0$ line of Fig. \ref{bhspace}, squeezed here into a point because
of the scale of this figure.

   The theoretical method to locate the future event horizon in Fig.
\ref{compbhglob} would now be to run a radial null geodesic backward in time
from the point $(\mu, \tau) = (\pi/2, \pi/2)$, i.e. from the image of the future
end of the AH+. However, for the most part, the AH+ runs so close to the crunch
singularity, and the geodesics intersecting the AH+ are so nearly tangent to
AH+, that numerical instabilities crash any such geodesic into the singularity
instantly, even if the initial point is chosen well away from $(\pi/2, \pi/2)$.
This happens all the way down to $\mu = 0.5$ at single precision and all the way
down to $\mu = 1.1$ at double precision. We did succeed, with double precision,
only at $\mu = 1.0$, and a null geodesic could be traced from there to the
center at $\mu = 0$. At the scale of Fig. \ref{compbhglob}, this whole geodesic
seems to coincide with the crunch and the AH+. However, it is well visible if
one closes in on the image of the area shown in Fig. \ref{bhspace}, the closeup
is shown in the inset.

   Actually, to make sure that we located the event horizon with an
acceptable precision, we ran three different null geodesics backward in time;
one from the point $\mu = 1.0, \tau = \tau_0 := \tau_{AH+}(1.0)$ right on the
apparent horizon, another one from $\mu = 1.0, \tau = \tau_1$, where $\tau_1$
was in the middle between the AH+ and the crunch singularity, and the third one
from $\mu = 1.0, \tau = \tau_2$, where $\tau_2$ was below the AH+, with the
time-difference $\tau_0 - \tau_2 = \tau_1 - \tau_0$. All three ran so close to
each other that they actually coalesced along the way and reached the center as
one curve; their coincidence in the inset in Fig. \ref{compbhglob} is thus not
an artefact of scale, but actual coincidence at double precision.


\begin{figure}[b]
 \includegraphics[scale = 0.45]{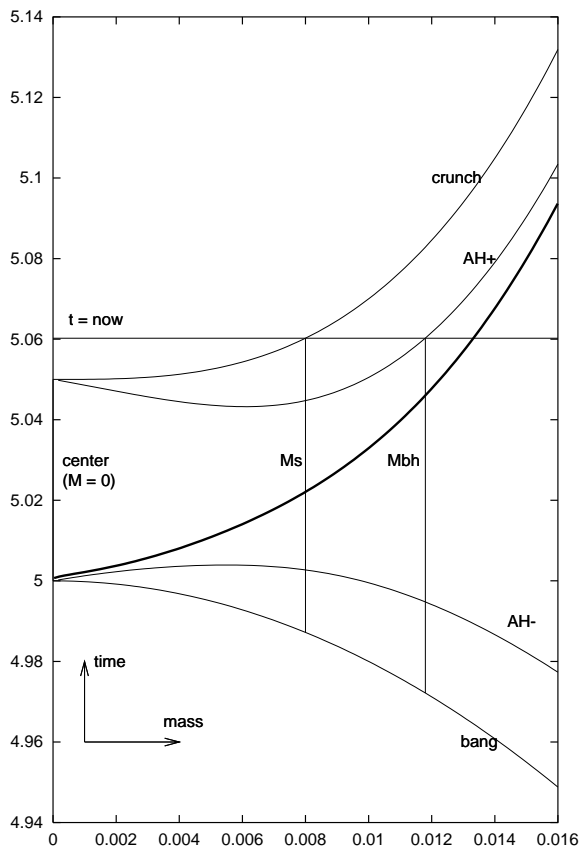}
 \caption{
 \label{evbhspace}
 The event horizon (thicker line) written into the frame of Fig.
\ref{bhspace}. Its intersection with the $M = 0$ axis does not coincide
with the central point of the Big Bang, this is only an illusion created
by the scale.}
 \end{figure}

   Finally, the event horizon had to be transformed back to the $(M, t)$-
coordinates and written into the frame of the figures \ref{bhspace} --
\ref{allgeomap}. This is done in Fig. \ref{evbhspace}. As stated earlier,
the event horizon is located between the geodesics no 5 and 6 from the
right in Fig. \ref{allgeomap}. By accident (caused by our choice of
numerical values in this example), the EH hits the center very close to
the central point of the Big Bang, but does not coincide with it.


   This whole construction should make it evident that there is no chance
to locate the event horizon by astronomical observations, even
approximately.  It only makes sense, in the observational context, to
speak about an upper limit on the mass inside the {\it apparent horizon}.
This is why we identified the observed mass of a black hole with the AH in
sec. \ref{design}%
 \footnote{Note that this model in which the event horizon has been found,
will actually only be used for the interior of our galaxy model -- i.e.
for that part of the galaxy that is invisible to outside observers because
of its proximity to the black hole horizon. The model of the visible part
of the galaxy will be different, and to locate the event horizon there,
the whole construction (integrating a null geodesic backward in time from
future null infinity) would have to be repeated. But, of course, the model
of a single galaxy does not extend to infinity. Hence, it hardly makes
sense to even speak of an event horizon in this context.}%
 .

 \section{The galaxy plus black hole formation model} \label{gala+bh}

   Our aim is to model the formation of a galaxy with a central black 
hole, starting from an initial fluctuation at recombination.  Our model 
consists of two parts joined together across a comoving boundary $M = 
M_{BH}$, with $M_{BH}$ the estimated present day mass inside the black 
hole horizon.  In the exterior part, we take existing observational data 
for the present day density profile, and the initial fluctuation is made 
compatible with CMB observations.  For the interior, no observational 
constraints exist, so we propose a couple of possible descriptions, as 
detailed below.  These are both LT models, and represent a collapsing 
body, and a dense 
 Kruskal-Szekeres wormhole in the sense of \cite{Hell1987}.

 \subsection{The black hole interior}
 \label{inmodel}

   Astronomical observations do not say anything about that portion of
galactic matter that had already fallen inside the apparent horizon by the
time the electromagnetic signal that would reach the observer was emitted.
What can be seen in the sky are only electromagnetic waves emitted by
objects that were still outside the AH at the time of emission.
Consequently, we are not constrained in any way in choosing a model for
the matter in the interior of a black hole, except for the need to match
it smoothly to a galaxy model.

   The term `black hole' is used in two disctinct ways.  Firstly, there is
the
 Schwarzschild-Kruskal-Szekeres black hole, which has the topology of two
universes joined by a temporary wormhole, and begins its life as a white
hole.  This was generalised to a
 matter-filled version in \cite{Hell1987}.  Secondly, there is the black
hole formed by the collapse of a massive body, which has an ordinary
topology without a wormhole.  Only in their late stages (after the closure
of the wormhole) do these two become essentially the same.  Both of these
can be reproduced by an
 L-T model, and we consider them in turn below.

   At this point, we must make a digression about the relation between
model black holes such as those considered here, and real collapsed
objects that are called ``black holes'' by astronomers. There are two
important points to be remembered:
 \begin{enumerate}

 \item   The matter proceeding toward a black hole disappears from the
field of view of any real observer before it hits the horizon, whatever
horizon is meant (see below). For example, the event horizon is the
boundary of the field of view for an observer at future null infinity,
i.e. one who is infinitely distant from the black hole and infinitely far
into the future. Therefore, the observationally determined ``mass of a
black hole'' is in fact only an upper limit of the mass that has actually
fallen within the horizon; the latter can never be measured in reality.
{\it Models} of black holes allow us to {\it calculate} better estimates
of that mass, and even if the arithmetic difference between a model
calculation and an observational limit is small, it is important to
understand the conceptual difference.

 \item   It is incorrect to speak about the event horizon in the context
of observations. We may know where the event horizon is only in a {\it
model}. In practice, we would have to be able to take into account the
future fate of every piece of matter, including those pieces that have
been outside our field of view up to now
 --- an obviously impossible task. Worse still, if the real Universe is to
recollapse in the future, we might already be inside the event horizon and
will never see any signature of it. Hence, the horizons whose signatures
we have any chance to see (like disappearance of matter from sight, or a
large mass being contained in a small volume) are {\it apparent horizons}
 --- they are local entities, detectable in principle at any instant
(although with the difficulty mentioned in point 1). Even in our simple model
considered in the previous section, it was rather difficult to determine the
position of the event horizon, and it could be done only numerically.

In the following we will identify the estimated black hole mass, obtained 
by fitting a model to present day observations, with $M_{BH}$, the mass 
within the apparent horizon at time $t_2$ (see eq. (\ref{msmbhat2})). 

 \end{enumerate}

 \subsection{A collapsed body}

   For this example of the interior, we use the model of section
\ref{IllustrationModel}, but with different values of the parameters.  We
will assume that $t_C(M)$ is an increasing function, to be able to create
a black hole, and to prevent shell crossings we assume $t'_B(M) < 0$ in
addition.  The functions have already been chosen so that the origin
conditions are satisfied.

   To assure a smooth match to the exterior model, we require the
continuity of the
 L-T arbitrary functions and their derivatives%
 \footnote{
 The Darmois junction conditions for matching an
 L-T model to itself across a comoving (constant $r$) surface only require
the matching of $E$ and $t_B$ at the same $M$.
 }
 at $M = M_{BH}$.  Given equations (\ref{exbafu}) and (\ref{exener}), we
solve for the constants $a$, $b$, $T$ \& $t_{B0}$, so that $E$, $t_B$,
$\tdil{E}{M}$ \& $\tdil{t_B}{M}$ are matched at the boundary:
 \begin{subequations}
   \label{EtBmatch}
 \begin{align}
   t_{B0} & = \left[ t_B - \frac{M}{2} \td{t_B}{M}
      \right]_{M = M_{BH}}, \\
   b & = - \left[ \frac{1}{2 M} \td{t_B}{M} \right]_{M = M_{BH}}, \\
   T_0 & = \left[ \frac{M}{6} \td{t_B}{M} + \frac{4 \pi M}{3 (-2 E)^{3/2}}
      - \frac{2 \pi M^2}{(-2 E)^{5/2}} \td{E}{M} \right]_{M = M_{BH}}, \\
   a & = \Bigg[\frac{1}{3M^2} \td{t_B}{M}
         + \frac{2 \pi} {3 M^2 (-2 E)^{3/2}} \nonumber \\
   &~~~~~~~~ + \frac{2 \pi} {M (-2 E)^{5/2}}\ \td{E}{M}
         \Bigg]_{M = M_{BH}}. 
 \end{align}
 \end{subequations}

 \subsection{A wormhole}
 \label{AWormhole}

   Since we have no way of knowing anything about the matter and spacetime
interior to $M_{BH}$, we can equally well fit in a
 dust-filled wormhole of the
 Kruskal-Szekeres type, constructed with the
 L-T metric.  The essential
requirement is that, at the middle of the wormhole, $M$ must have a minimum
value $M_{\text{min}}$, and $E(M_{\text{min}}) = -1/2$.  The minimum lifetime
(time from past to future singularity) of the wormhole is then $2 \pi
M_{\text{min}}$.  We choose the following functions:
 \begin{subequations}
   \label{WHEtB}
 \begin{align}
   t_B & = t_{B0} - b (M - M_{\text{min}})^2 , \\
   E & = \frac{-M_{\text{min}}}{2 M} + a (M - M_{\text{min}}) .
 \end{align}
 \end{subequations}
 From these, the conditions for matching to an exterior at some given $M$
value are
 \begin{widetext}
 \begin{subequations}
   \label{EtBWHmatch}
 \begin{align}
   M_{\text{min}} &= \Bigg[ M^2 \td{E}{M} + M \Bigg( 1 
      \pm \sqrt{1 + 2 E + M^2 \left( \td{E}{M} \right)^2}\; 
      \Bigg) \Bigg]_{M = M_{BH}}, \\
   a &= \Bigg[ \frac{1}{2} \td{E}{M} - \frac{1}{2 M} \Bigg( 1
      \pm \sqrt{1 + 2 E + M^2 \left( \td{E}{M} \right)^2}\; \Bigg)
      \Bigg]_{M = M_{BH}}, \\
   b &= \Bigg[ \td{t_B}{M} \Bigg/ \Bigg\{ 2 M \Bigg( M \td{E}{M}
      \pm \sqrt{1 + 2 E + M^2 \left( \td{E}{M} \right)^2}\; \Bigg)
      \Bigg\} \Bigg]_{M = M_{BH}},  \\
   t_{B0} &= \Bigg[ t_B + \frac{M}{2} \td{t_B}{M} \Bigg( M \td{E}{M}
      \pm \sqrt{1 + 2 E + M^2 \left( \td{E}{M} \right)^2}\;  \Bigg)
      \Bigg]_{M = M_{BH}}.
 \end{align}
 \end{subequations}
 \begin{figure}[h]
 \includegraphics[scale = 0.41]{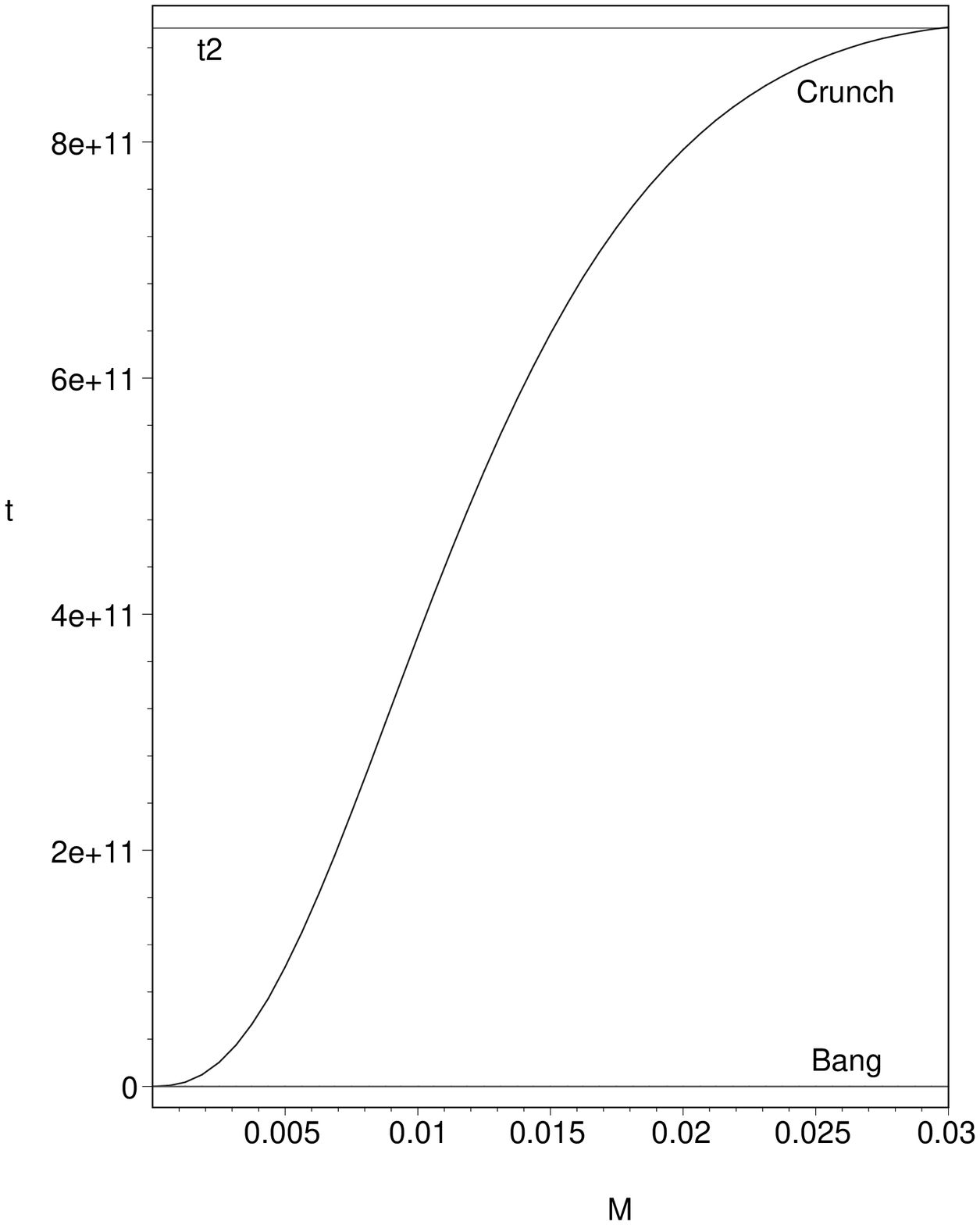}
 \hspace*{10mm}
 \includegraphics[scale = 0.41]{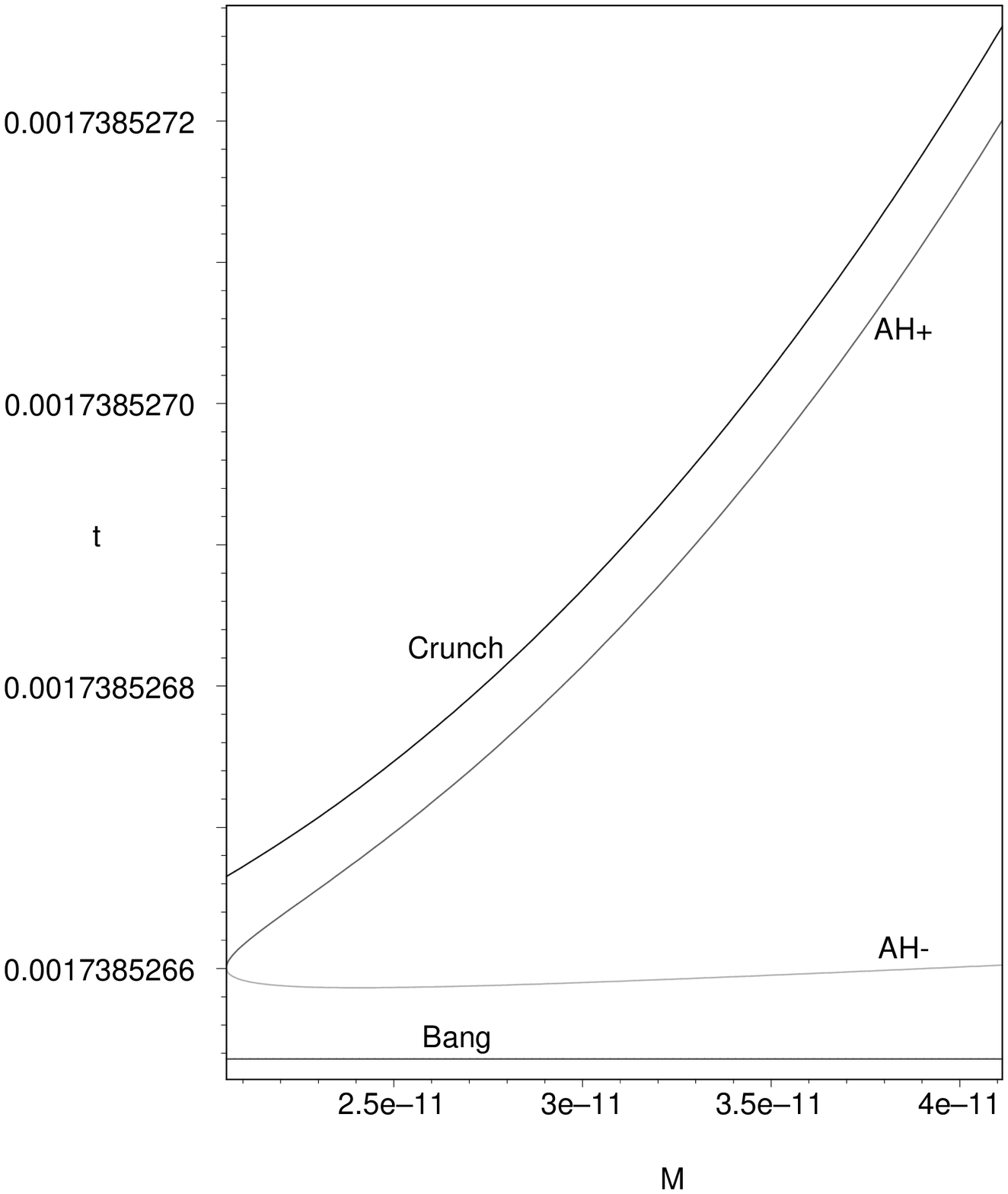}
 \caption{
 \label{WmHl}
 The
 L-T model for the wormhole interior of section \ref{AWormhole} with
parameters (\ref{MatchedWmHlParams}), in geometric units.  The second
graph shows the neck of the wormhole magnified about $10^9$ times.  Only
on this scale are the apparent horizons distinguishable from the bang \&
crunch.  The bang does actually curve downwards, but its variation is much
less than that of the crunch.  The neck is at the left where the mass
reaches a minimum ($M_{\text{min}} = M_0$).  The apparent horizons cross
in the neck at the moment of maximum expansion.  The second sheet is not
shown.  It could be a mirror image of this sheet, or it could be quite
different.  The boundary with the exterior model (at $M_{BH} = 0.03$) is
defined to be where $t_2 = t_{\text{AH}}$.  Recombination $t_1$ is
indistinguishable from the bang in the first graph, and far to the future
in the second.}
 \end{figure}
 \end{widetext}
 This model was chosen for simplicity, and so is not very flexible.  The
matching fixes the value of $M_{\text{min}}$, which determines the
lifetime of the wormhole.  A model with more parameters would allow the
wormhole lifetime to be a free parameter. The apparent horizons and
singularities for this model with the parameters (\ref{MatchedWmHlParams})
used below are illustrated in Fig. \ref{WmHl}.
 %
 %

 \subsection{The exterior galaxy model}
 \label{design}

 \subsubsection{The final density profile}

\begin{figure}[b]
 \includegraphics[scale = 0.45]{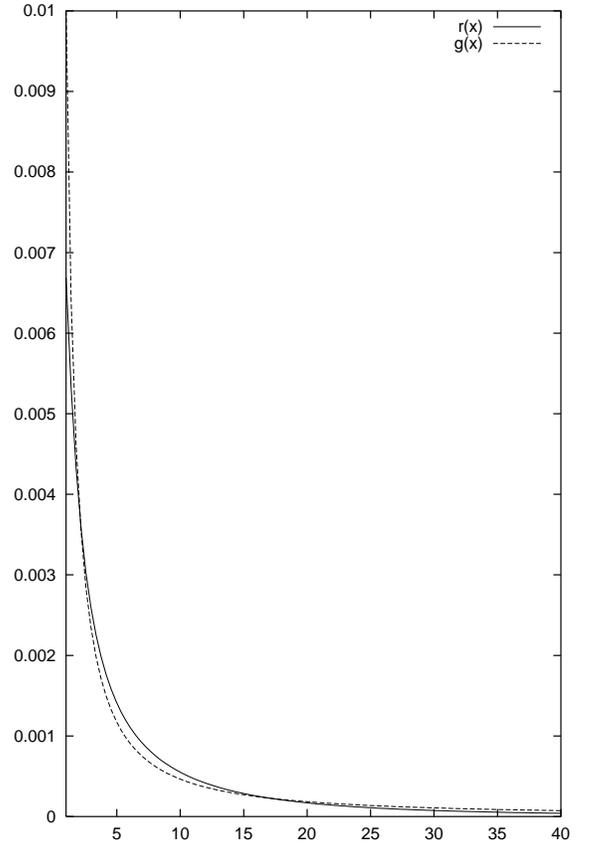}
 \caption{
 \label{galprof}
 Comparison of the profile (\ref{m87rpro}) (marked as $r(x)$) with
(\ref{myrprof}) (marked as $g(x)$).  The $\rho$ on the vertical axis is
mass density in units of $10^{-23}$ g/cm$^3$, the $x$ on the horizontal
axis is the distance from the center in arc minutes, just as in Ref.
\cite{FaLG1980}. The values of parameters are given below eq.
(\ref{m87rpro}).}
 \end{figure}

   As our example for the density profile of the final state, we choose
the galaxy M87. It is believed to contain a large black hole
\cite{Georww}, and the density profile for its outer part had been
proposed some time ago \cite{FaLG1980}:
 \begin{equation}\label{m87rpro}
   \rho(s) = \rho_0\left/\left(1 + bs^2 + cs^4 + ds^6\right)^n\right.,
 \end{equation}
 where $\rho_0 = 1.0\cdot 10^{-25}$ g/cm$^3$, $b = 0.9724$, $c =
3.810\cdot 10^{-3}$, $d = 2.753\cdot 10^{-8}$, $n = 0.59$. The distance
from the center, $s$, is measured in arcmin, i.e. it is dimensionless, and
is related to the actual distance $r$ by
 \begin{equation}\label{arcvsdis}
   s = \frac r D \frac {21600} {\pi} := r\delta,
 \end{equation}
 where $D$ is the distance from the Sun to the galaxy. (Of course, no
galaxy and no real black hole is spherically symmetric, so we cannot model
any actual galaxy with the L-T solution. However, we wish to make our
illustrative example as close to reality as possible, and this is why we
stick to an actual object.) For our purposes, we need density expressed as
a function of mass, and a profile that goes to infinity at $r \to 0$, to
allow for the singularity inside the black hole%
 \footnote{
 Even though we will use the interior model near the centre, this
requirement assists in joining interior and exterior smoothly.
 }%
 .  The mass profile corresponding to (\ref{m87rpro}) is not an
elementary function. However, it turns out that the following very simple
profile is a close approximation to (\ref{arcvsdis}) in the region considered in
Ref. \cite{FaLG1980} (see Fig. \ref{galprof}):
 \begin{equation}\label{myrprof}
   \rho(r) = \rho_0/(\delta r)^{4/3},
 \end{equation}
 with the same value of $\rho_0$.
 %
 %
 The corresponding mass distribution is
 \begin{equation}\label{premassvsr}
   \widetilde{M}(r) = \frac {12}5 \pi \rho_0 r^{5/3}\delta^{-4/3},
 \end{equation}
 and we make it singular at $r = 0$ by adding a constant $M_S$ to
$\widetilde{M}(r)$ (this is the same $M_S$ as defined before
 --- the mass that had already fallen into the singularity at $t_2$; so
far this is still an arbitrary constant):
 \begin{equation}\label{massvsr}
   M(r) = M_S + \frac {12}5\pi \rho_0 r^{5/3}\delta^{-4/3}.
 \end{equation}
 Hence, $r = [5(M - M_S)\delta^{4/3}/12\pi \rho_0]^{3/5}$, and
 \begin{equation}\label{rhovsm}
   \rho(M) = (12\pi/5)^{4/5}{\rho_0}^{9/5}\delta^{-12/5}(M - M_S)^{-4/5}.
 \end{equation}
 In principle, the value of $\rho$ near to $M = M_{BH}$ should be
measureable, but in this regime the difference between the Newtonian and
 L-T definitions of density becomes too pronounced (because of the
 non-flat geometry). To infer $\rho(M_{BH})$ in a sensible way, the
results of observations should be consistently reinterpreted within the
 L-T scheme, and such results are not, and will not be available for a
long time. Consequently, we will have to give up on this bit of
information.

   From (\ref{rhoRM}), we find the corresponding $R(M)$ via $R^3 =
(3/4\pi)\int_{M_S}^M \d x/\rho(x)$, which is
 \begin{equation}\label{rvsm}
   R(M) = \left( \frac 5 {12\pi\rho_0}\right)^{3/5} \delta^{4/5}
   \left(M -M_S\right)^{3/5}.
 \end{equation}
 Now we can determine the constant $M_S$ by the requirement that $R = 2M$
at $M = M_{BH}$, i.e.
 \begin{equation}\label{eqforms}
   \left( \frac 5 {12\pi\rho_0}\right)^{3/5} \delta^{4/5}
   \left(M_{BH} -M_S\right)^{3/5} = 2M_{BH}.
 \end{equation}
 From here we find
 \begin{equation}\label{defms}
   M_S = M_{BH} - \frac {12\pi\rho_0} {5\delta^{4/3}}
   \left(2M_{BH}\right)^{5/3}.
 \end{equation}
 It follows, as it should, that $M_S < M_{BH}$. However, this result makes
sense only if the $M_S$ thus defined is positive. The condition $M_S > 0$
is equivalent to
 \begin{equation}\label{posmscond}
   \left(\frac {24}5 \pi \rho_0\right)^{3/2}\  \cdot \
   \frac {2M_{BH}} {\delta^2} < 1.
 \end{equation}
 For checking this inequality, all quantities have to be expressed in
geometric units. The black hole in M87 is believed to have mass $M_{BH} =
3\cdot 10^9 M_{\odot}$ \cite{MAFW2002,MacMarAxo97}, and its distance from
the Sun is $D = 43 \cdot 10^6$ light years. In geometric units, with 1
year = 31 557 600 s, $c = 3\cdot 10^9$ cm/s, $G = 6.6726\cdot 10^{-
8}\text{cm}^3/\text{g}\cdot \text{s}^2$ and $M_{\odot} = 1.989\cdot
10^{33}$ g, this makes $M_{BH} = 4.424\cdot 10^{14}$cm, $\delta =
5.067\cdot 10^{-22} \text{cm}^{-1}$, $\rho_0 = 0.741 \cdot 10^{-53}
\text{cm}^{-2}$, and the left-hand side of (\ref{posmscond}) comes out to
be $4.075\cdot 10^{-21}$, which is very safely within the limit.

 \subsubsection{The initial fluctuation}
 \label{InitExtFluc}

   In order to define a model uniquely, we only need one more profile for
the region $M \geq M_{BH}$, e.g. a density or velocity profile at $t = t_1
< t_2$ or a specific choice of $E(M)$ or $t_B(M)$.  Apart from the density
profile at $t = t_2$, there are no other observational constraints in the
region $M \geq M_{BH}$
 --- the time by which galaxies started forming is not well known, and
presumably different for each galaxy, nothing is known about the initial
density or velocity distribution in the
 proto-galaxy at that time.

   Since the only quantities that are to some degree constrained by the
observations are the density and velocity profiles at the recombination epoch,
it will be most natural to use these for $t_1$.  Even so there is a problem: no
numerical data are available for amplitudes of the temperature fluctuations of
the CMB radiation at such small scales. We consequently chose a zero velocity
fluctuation for one case and a zero density fluctuation for another. However it
turned out that the latter was not suitable, since the solution required a
collapsing hyperbolic region near $M_{BH}$ in the exterior model.

 \section{Numerical evolution of the models}

   The programs written for papers I \& II were adapted to facilitate this
two-step model construction.  First the exterior profiles were used as input to
solve numerically for $E(M)$ and $t_B(M)$ for $M_{BH} \leq M \leq 1$.  The
values of $E$ \& $t_B$ and their derivatives at $M_{BH}$ were extracted, and the
parameters of the interior model calculated from them. Then the functions $E$ \&
$t_B$ were numerically extended into the interior model, down to $M = 0$ or $M =
M_{\text{min}}$.  From this data, the model evolution was reconstructed using
existing programs.

\begin{figure*}[!]
  \includegraphics[scale = 0.4]{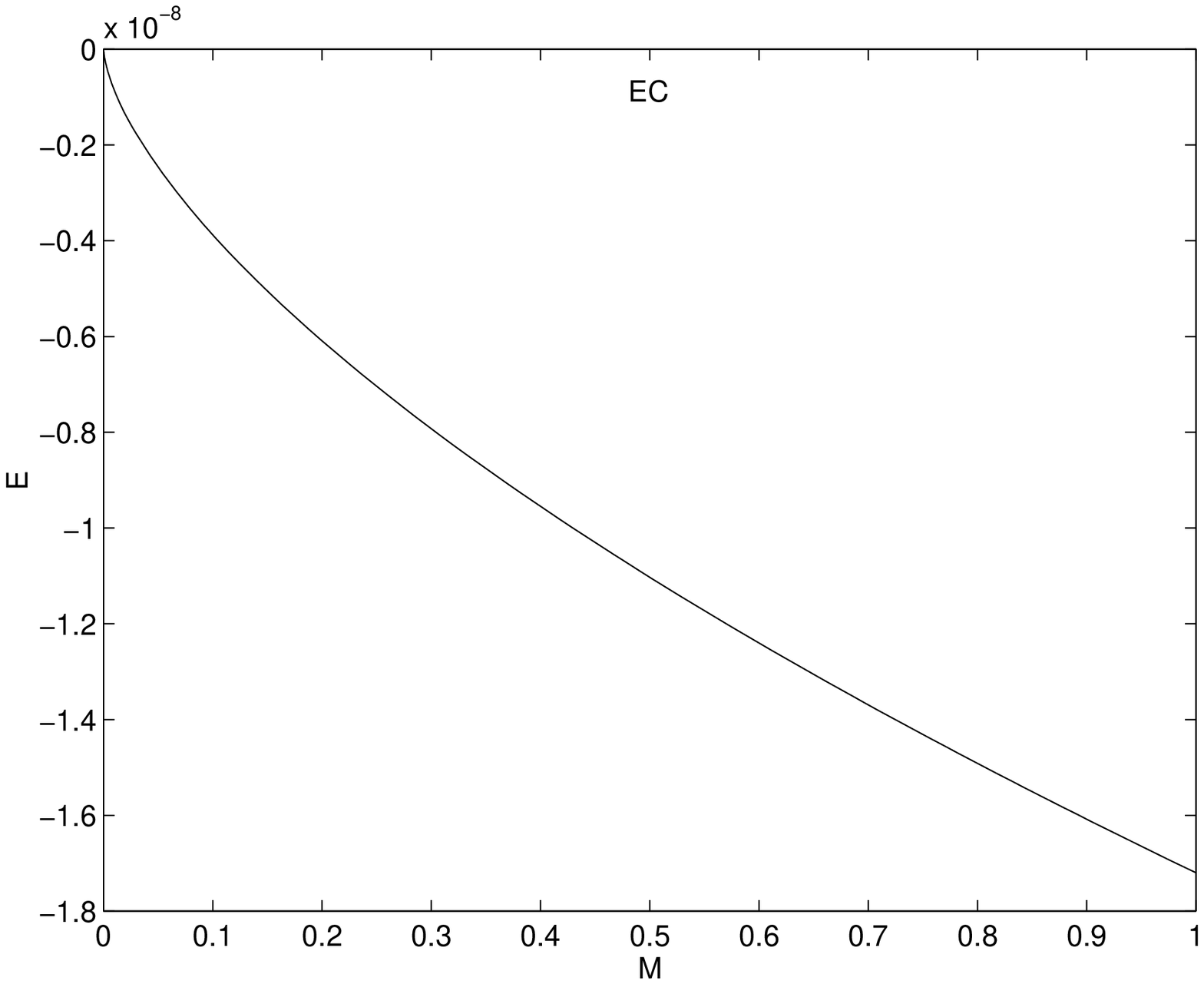}
  \hspace*{10mm}
  \includegraphics[scale = 0.4]{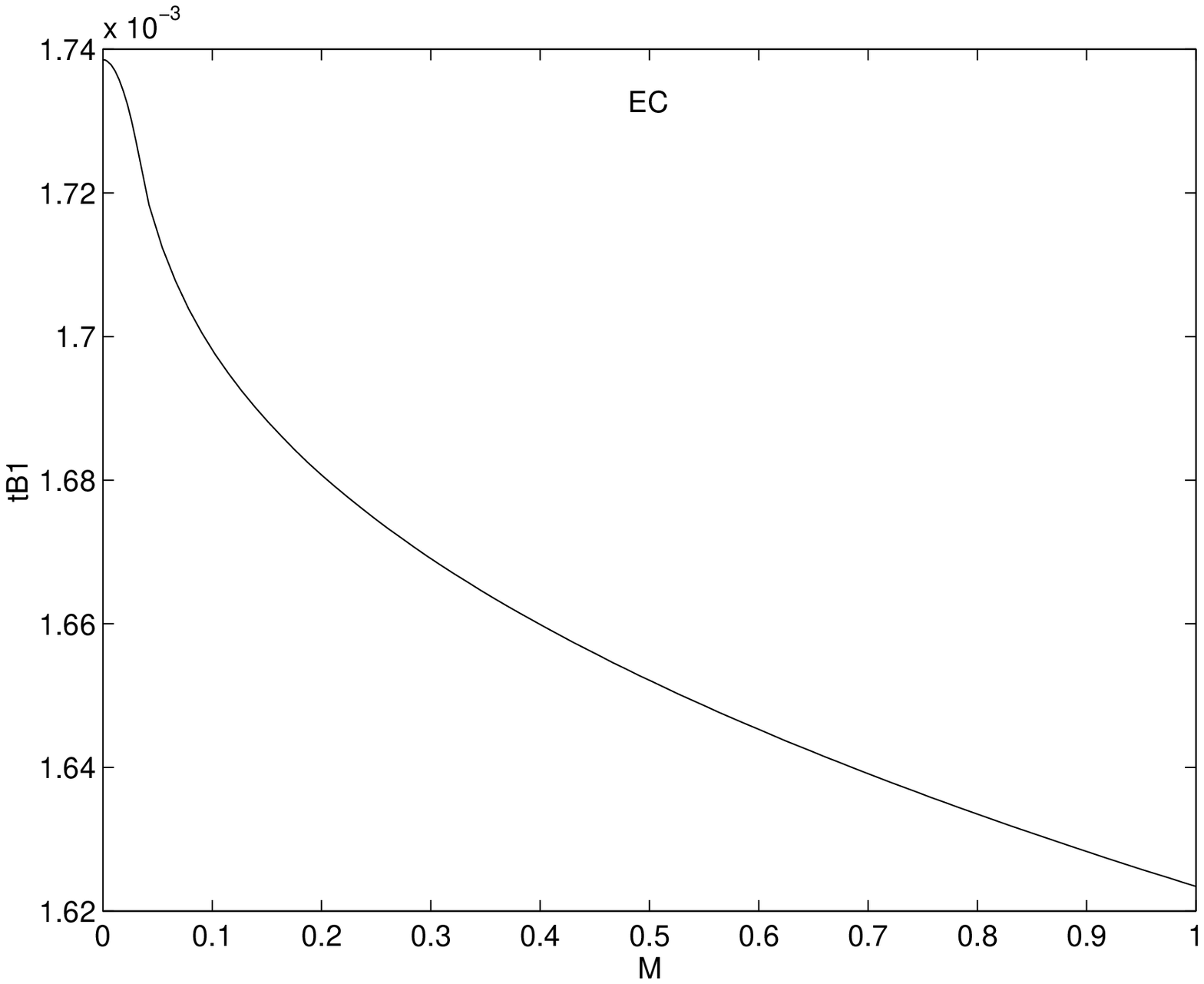}
  \\[5mm]
  \includegraphics[scale = 0.4]{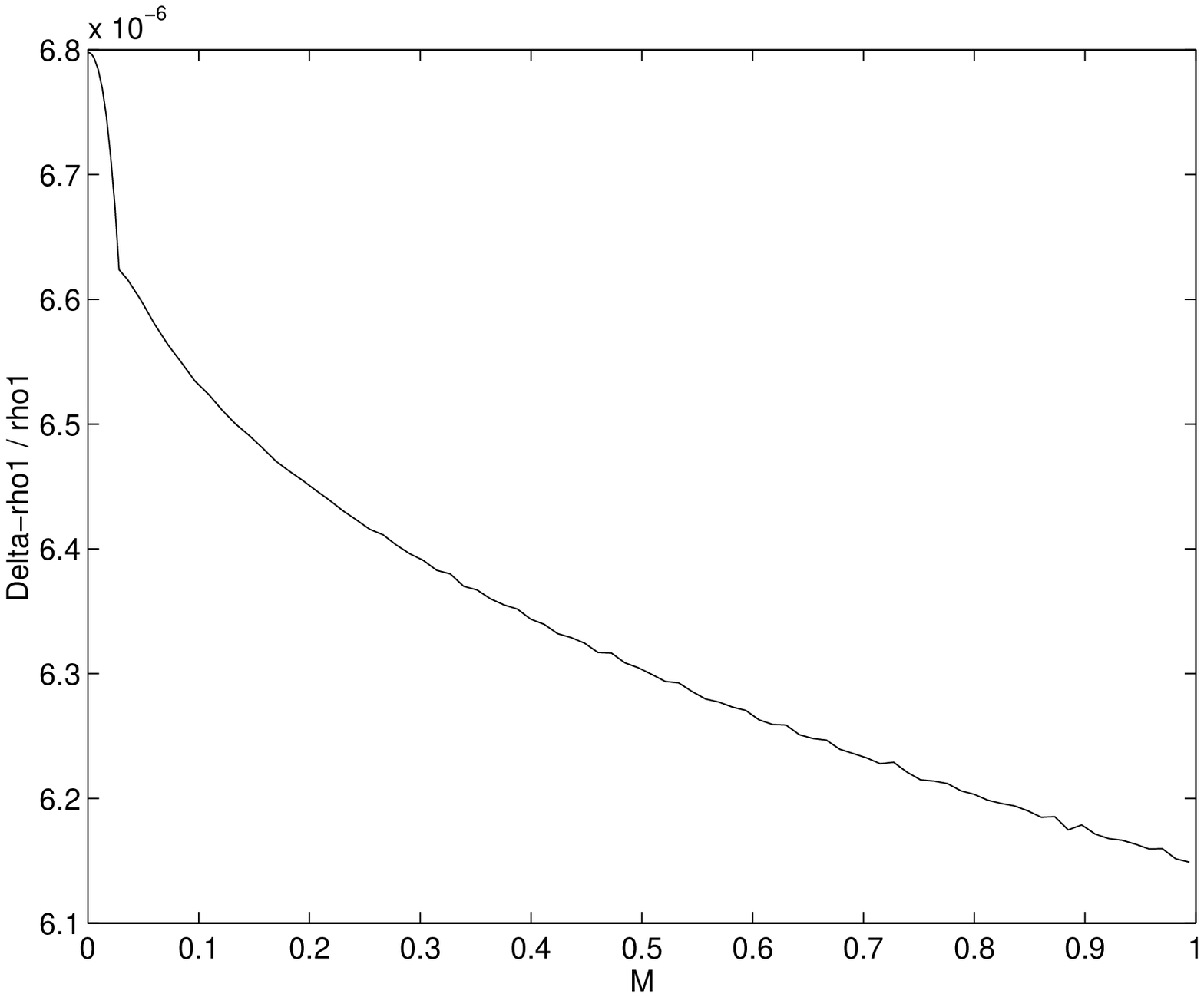}
  \hspace*{10mm}
  \includegraphics[scale = 0.4]{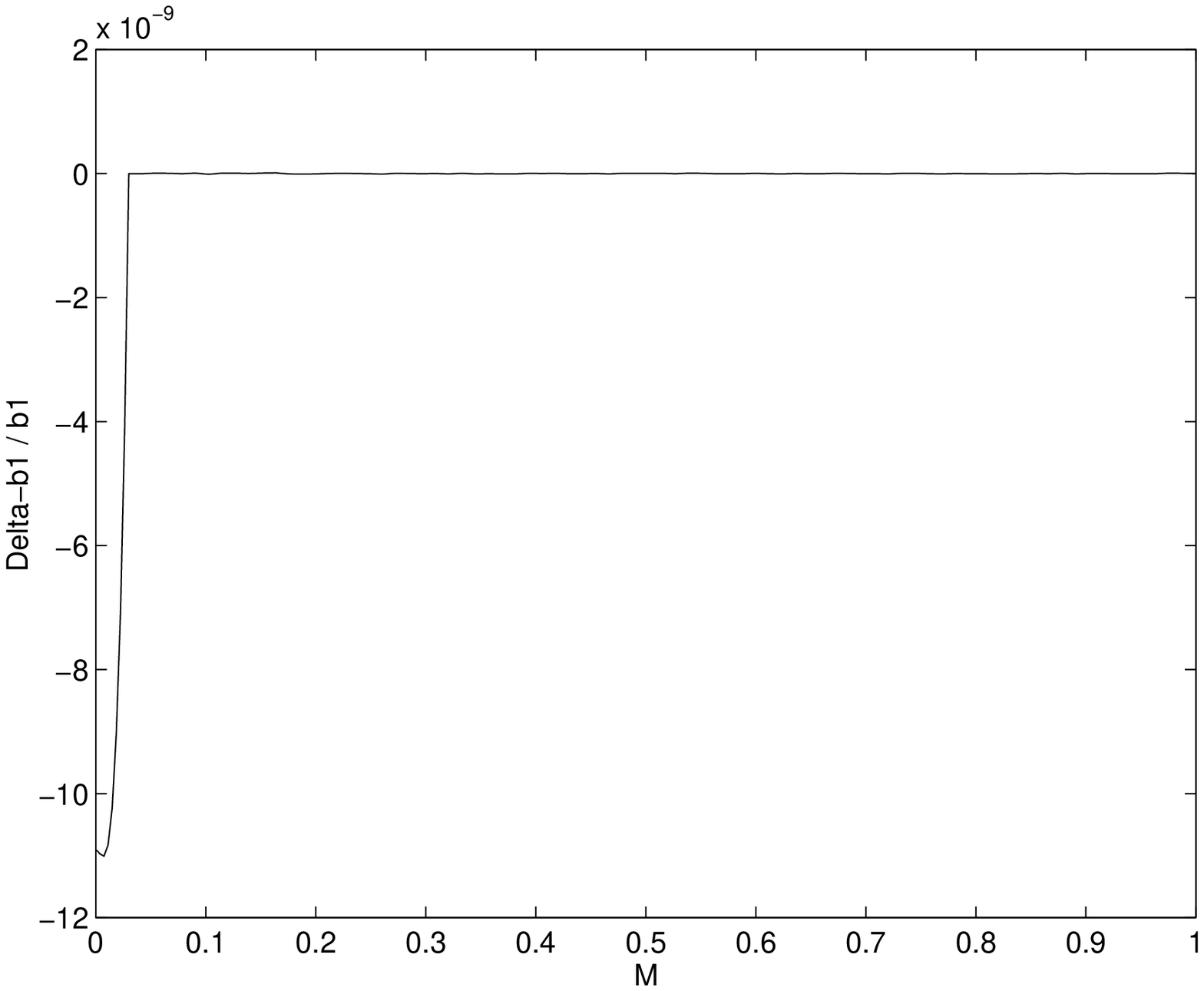}
  \\[5mm]
  \includegraphics[scale = 0.4]{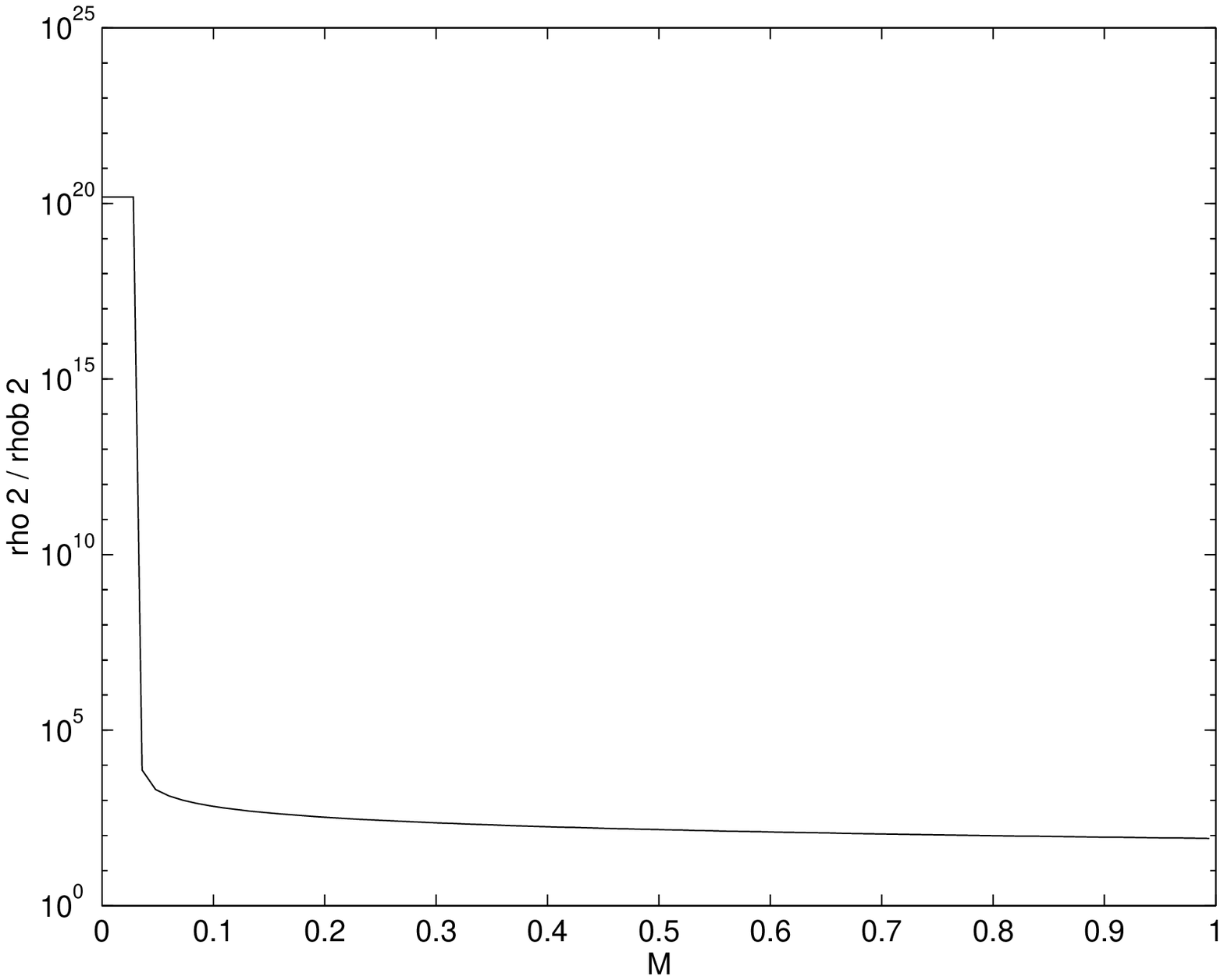}
  \hspace*{10mm}
  \includegraphics[scale = 0.4]{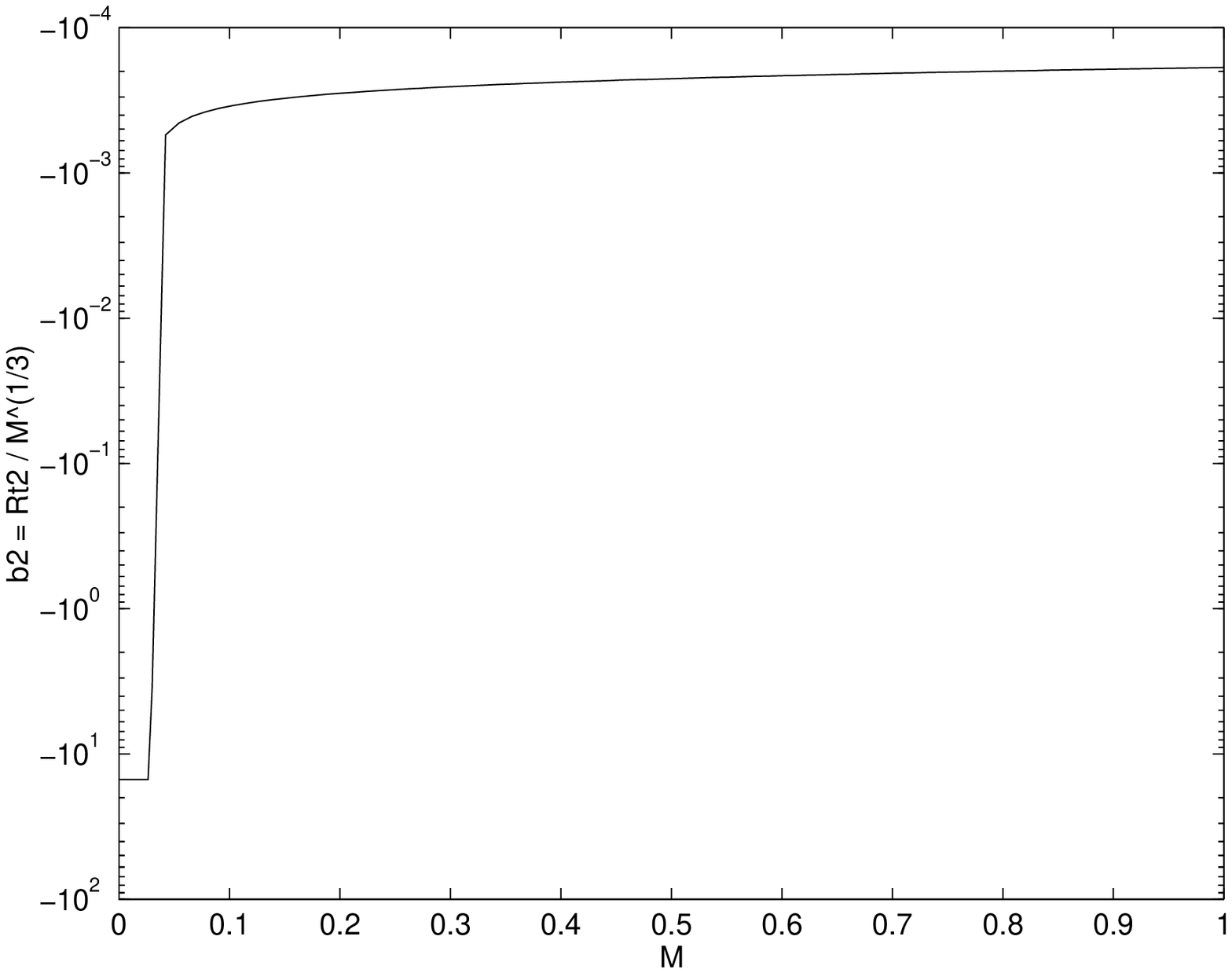}
 \caption{
 \label{V0D30figA}
 The
 L-T model for the formation of a galaxy that develops a central black
hole. Shown are the
 model-defining
 L-T functions $E(M)$ \& $t_B(M)$, the $\rho_1(M)$ and $b_1(M)$
fluctuations, the $\rho_2(M)$ and $b_2(M)$ variations.  The $b_2(M)$
variation is zero, and only very small numerical error shows.  Note that
the graphs have been clipped at $\log(R) = 0$, $\log(b) = 15$ and
$\log(\rho) = -5$ (geometric units).  The ``EC" indicates the range
considered is an elliptic region that is collapsing by $t_2$.}
 \end{figure*}

\begin{figure*}[!]
  \includegraphics[scale = 0.45]{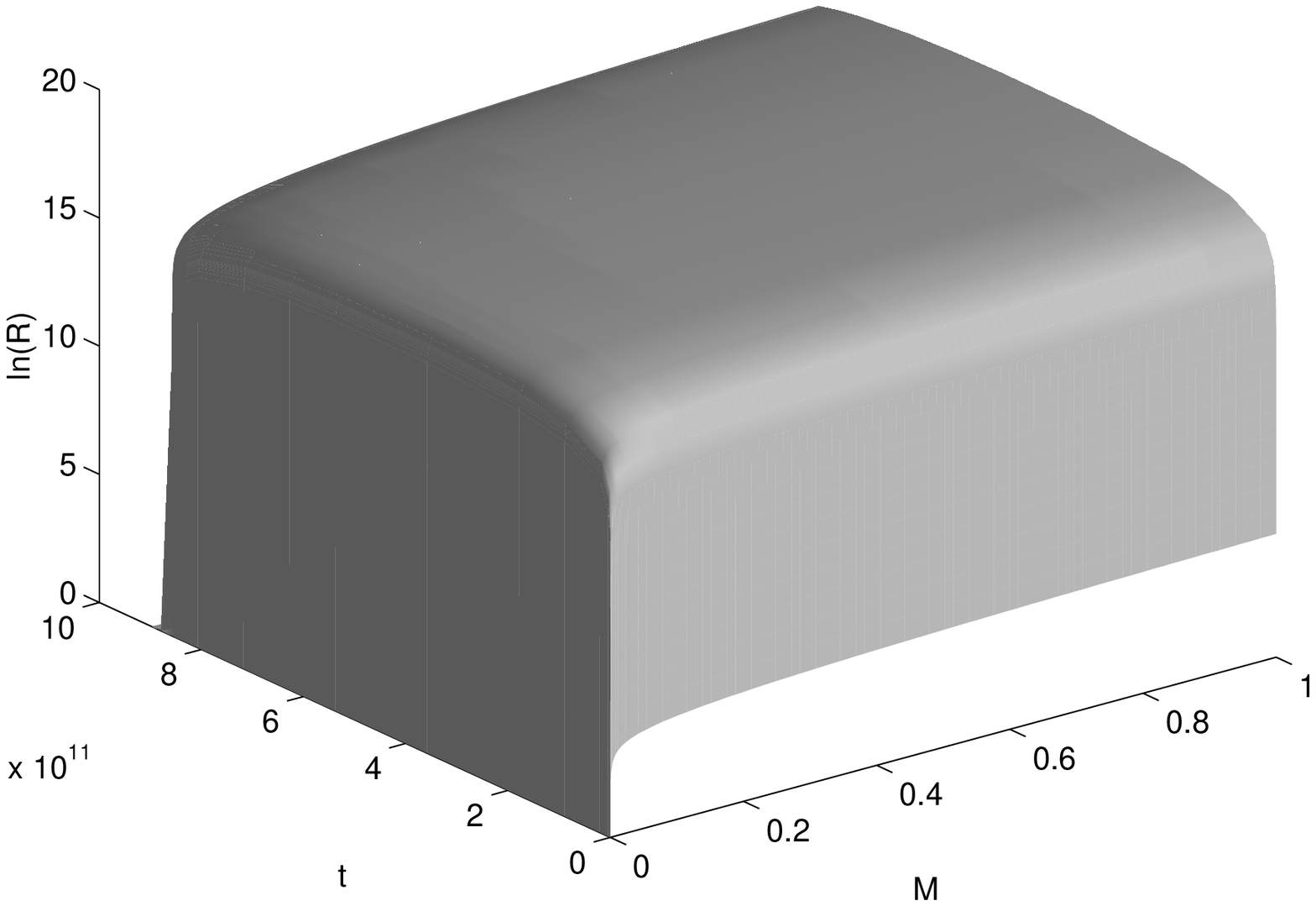}
  \hspace*{10mm}
  \includegraphics[scale = 0.45]{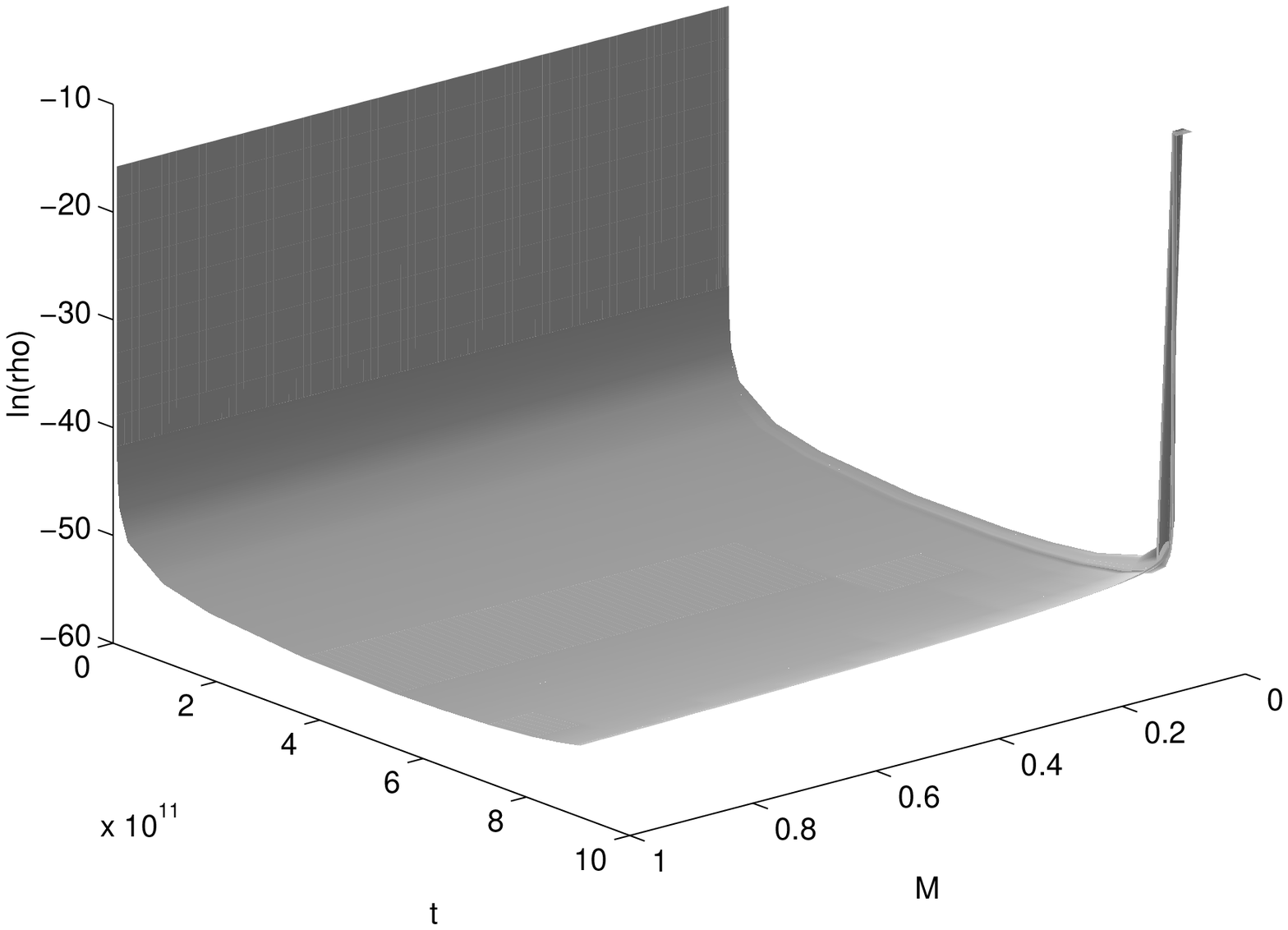}
 \caption{ \label{V0D30figB}
 The evolution of the
 L-T model for the formation of a galaxy that develops a central black
hole. Shown are the evolution of $R(t,M)$ and of $\rho(t,M)$ for run
Vi0$\rho$f30.  In the $R(t,M)$ graph, the origin $R(t, 0) = 0$ is on the
left, and expansion to a maximum occurs as time increases towards the
``north-west".  At the inital time, $t_1 =$ recombination, $R>0$ except at
the origin.  In the $\rho(t,M)$ graph the view has been rotated by
180$^\circ$ relative to the $R(t,M)$ graph for clarity, so that the orgin
is on the right and time increases towards the ``south-east".  By $t_2 =$
today, the innermost region has collapsed to a black hole of mass $3
\times 10^9~M_\odot$.  Note that the graphs have been clipped at $\log(R)
= 0$ and $\log(\rho) = -5$ (geometric units).}
 \end{figure*}

   Our first model uses the final density profile of section \ref{design}
for the galaxy at time $t_2 = 14$~Gyr, and a flat initial velocity profile
at time $t_1 = 10^5$~y, both exterior to $M_{BH}$.  The interior of
$M_{BH}$ is a black hole formed by collapse, as described by
(\ref{exbafu}) and (\ref{exener}), with parameters determined by the
matching (\ref{EtBmatch}).  Geometric units were chosen such that
$10^{11}~M_\odot$ is the mass unit.  In these units, the parameters are:
 \begin{multline}\label{MatchedClBdParams}
   a = 9.0662 \times 10^{14} ~,~~~~
   b = 0.012409 ~, \\
   t_{B0} = 0.0017385 ~,~~~~
   T_0 = 8.7279 \times 10^{11}
 \end{multline}
 The resulting arbitrary functions and the behaviour of the combined model
are shown in Figs \ref{V0D30figA} and \ref{V0D30figB}.  Notice that the
fluctuations of both density and velocity at recombination are well within
$3 \times 10^{-5}$ and $10^{-4}$.  The black hole singularity forms at
time $T_0 = 13.618$~Gyr (since $t_{B0} = 2.7126 \times 10^{-5}$~y is
negligible), so it is 400 million years old by today.
 %

\begin{figure*}[!]
  \includegraphics[scale = 0.4]{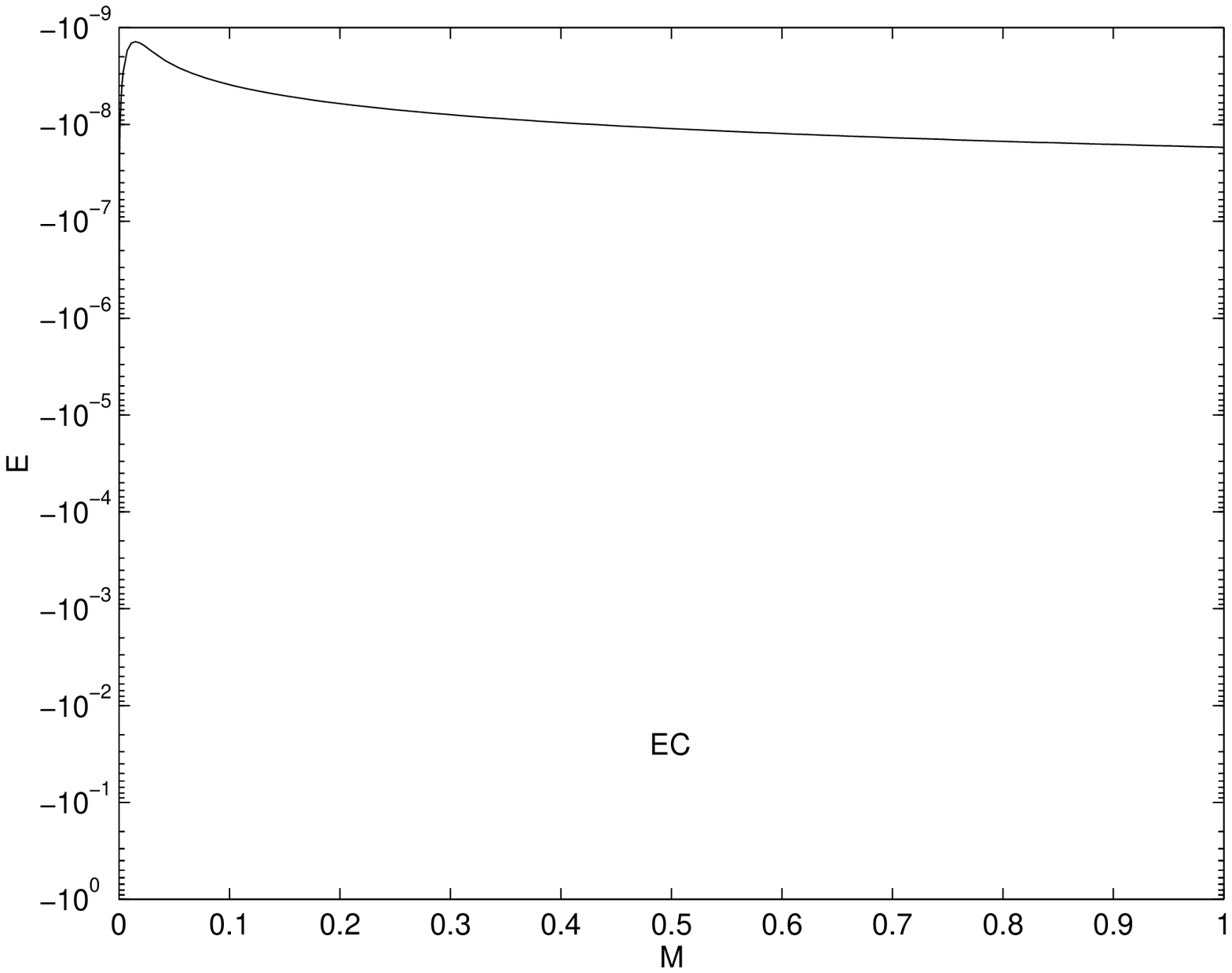}
  \hspace*{10mm}
  \includegraphics[scale = 0.4]{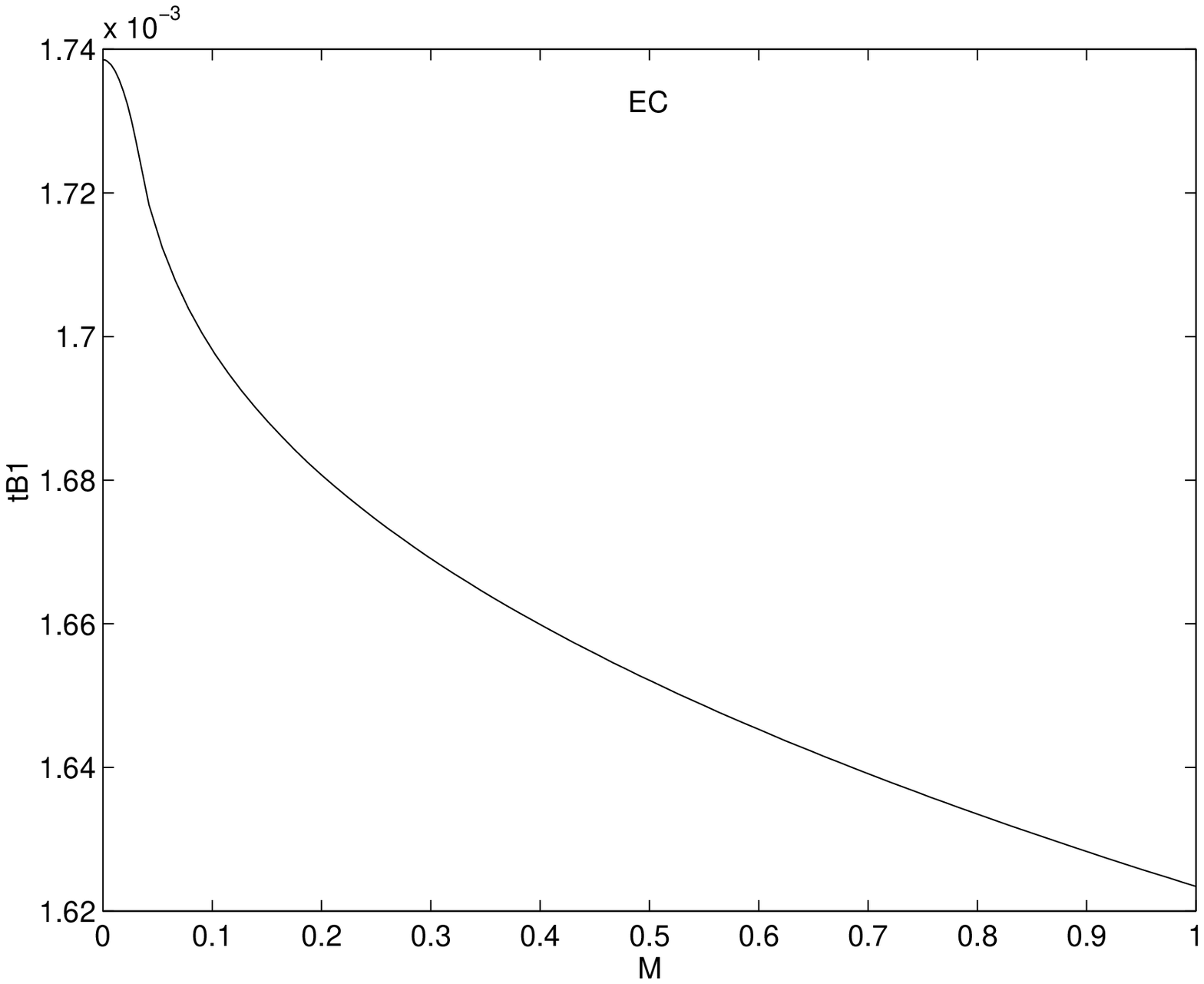}
  ${}$ \\[5mm]
  \includegraphics[scale = 0.4]{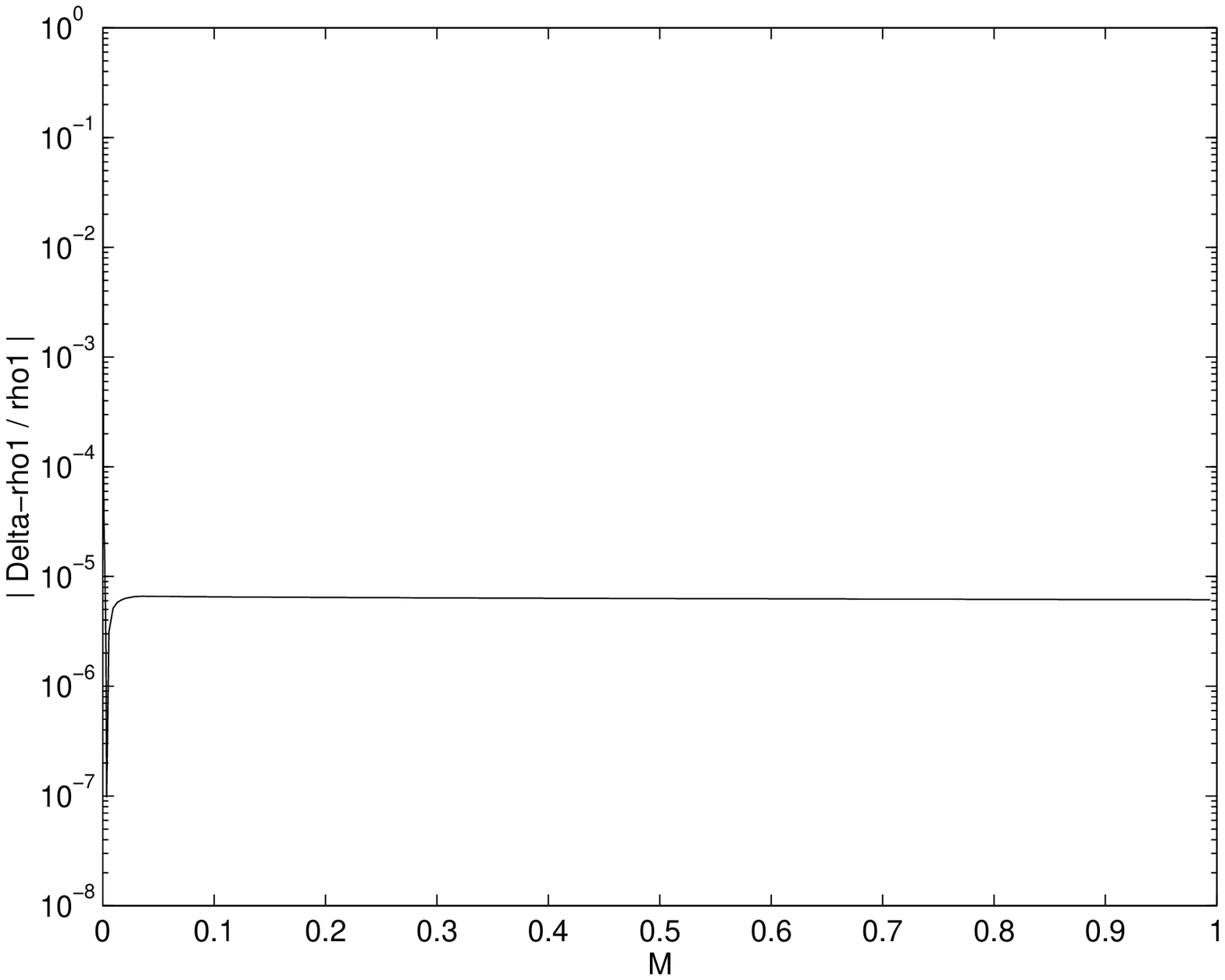}
  \hspace*{10mm}
  \includegraphics[scale = 0.4]{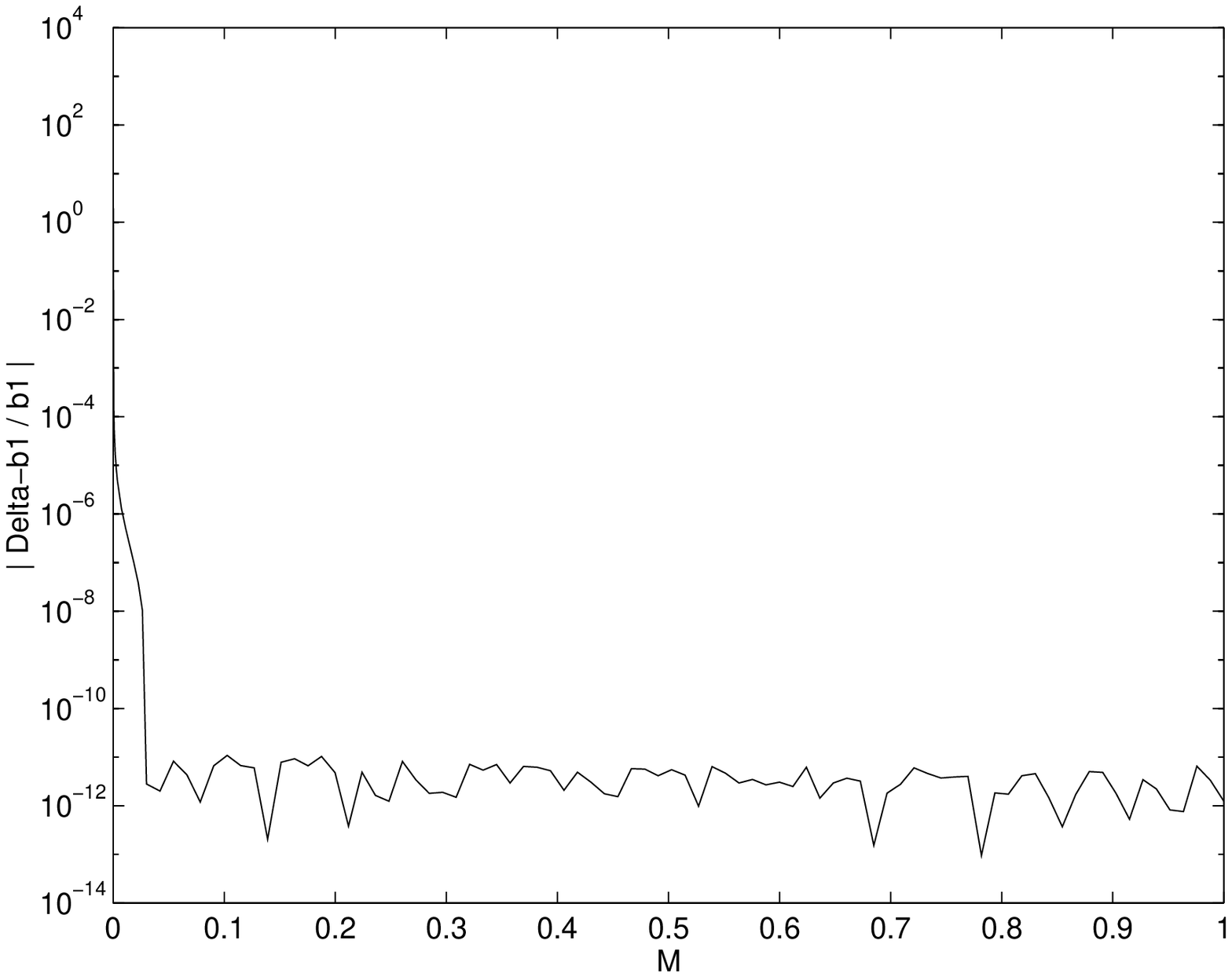}
  ${}$ \\[5mm]
  \includegraphics[scale = 0.4]{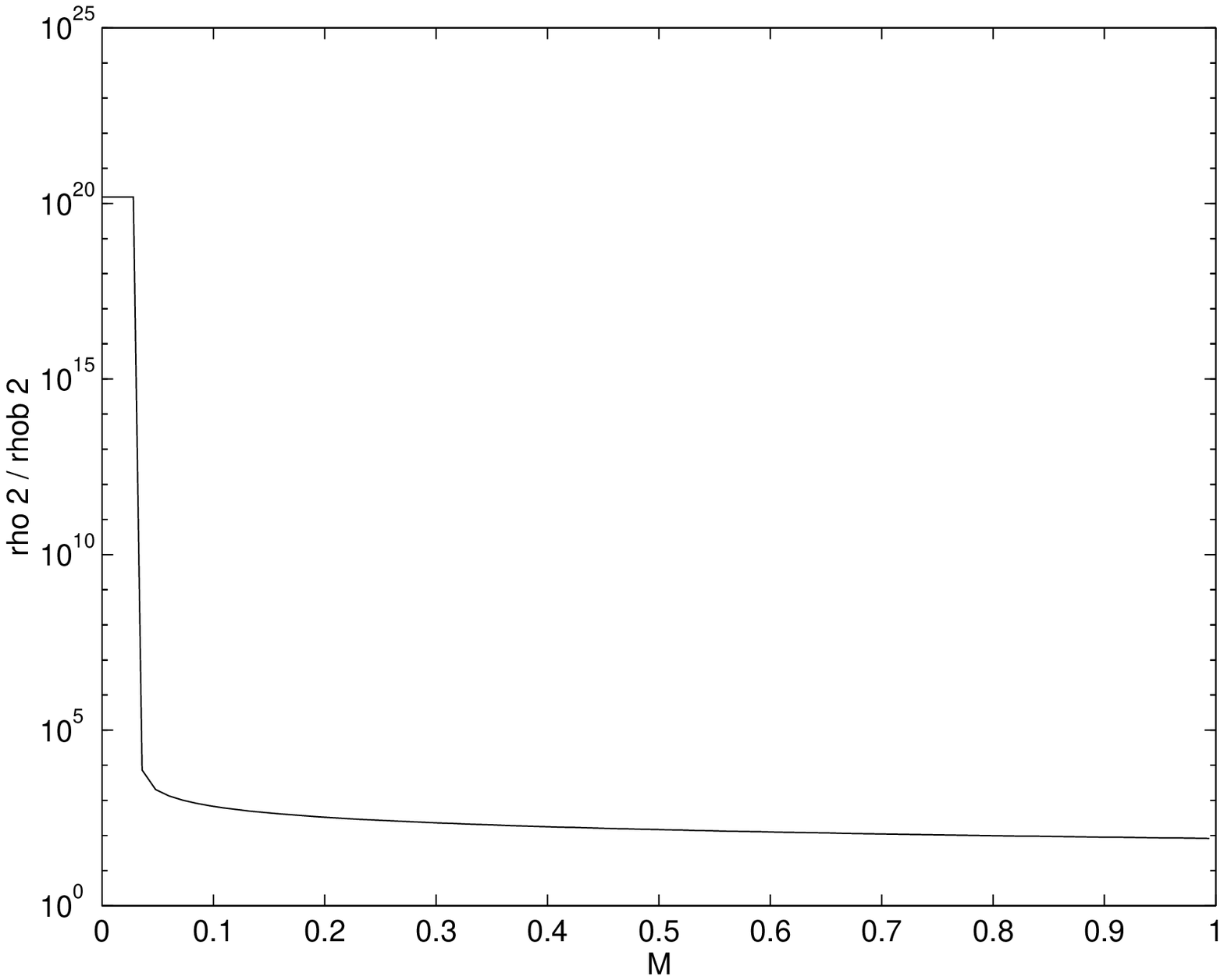}
  \hspace*{10mm}
  \includegraphics[scale = 0.4]{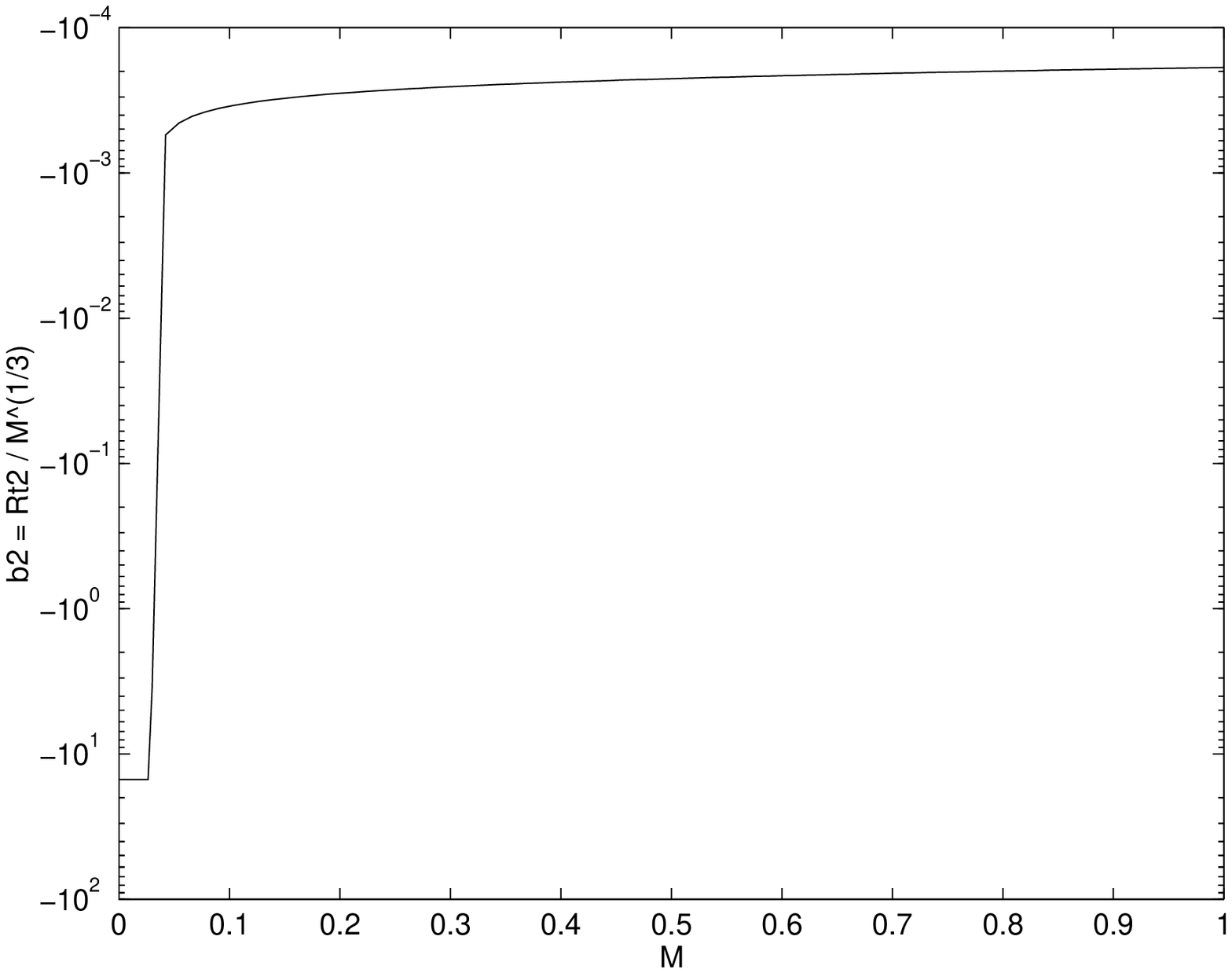}
 \caption{ \label{WHV0D30figA}
 The
 L-T model for the formation of a galaxy around a
 pre-existing central black hole.  Shown are the
 model-defining
 L-T functions $E(M)$ \& $t_B(M)$, the $\rho_1(M)$ and $b_1(M)$
fluctuations, the $\rho_2(M)$ and $b_2(M)$ variations.  (Again the
$b_2(M)$ variation is pure numerical error.)  Note that the graphs have
been clipped at $\log(R) = 0$, $\log(b) = 15$ and $\log(\rho) = -5$
(geometric units).}
 \end{figure*}

\begin{figure*}[!]
  \includegraphics[scale = 0.45]{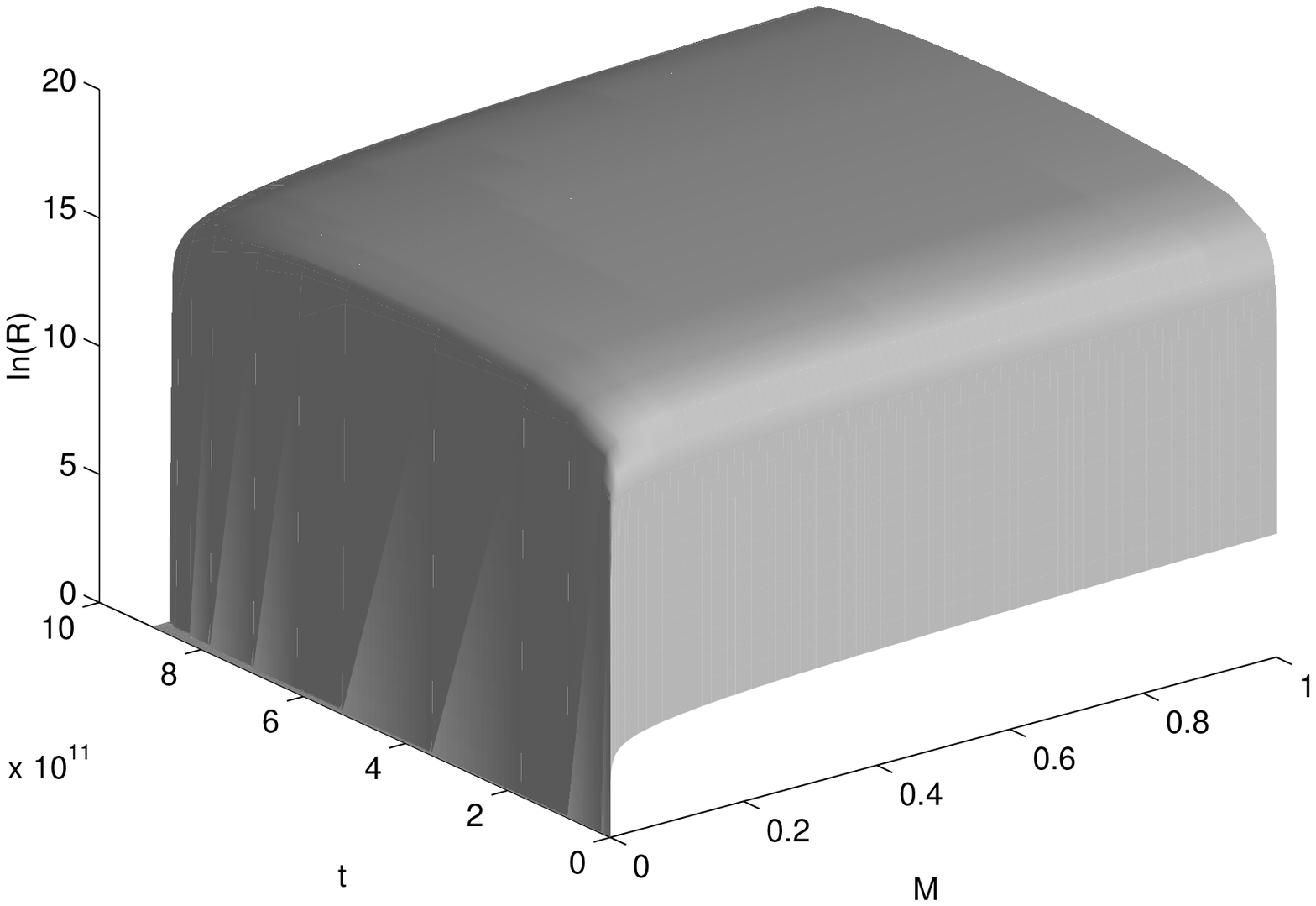}
  \hspace*{10mm}
  \includegraphics[scale = 0.45]{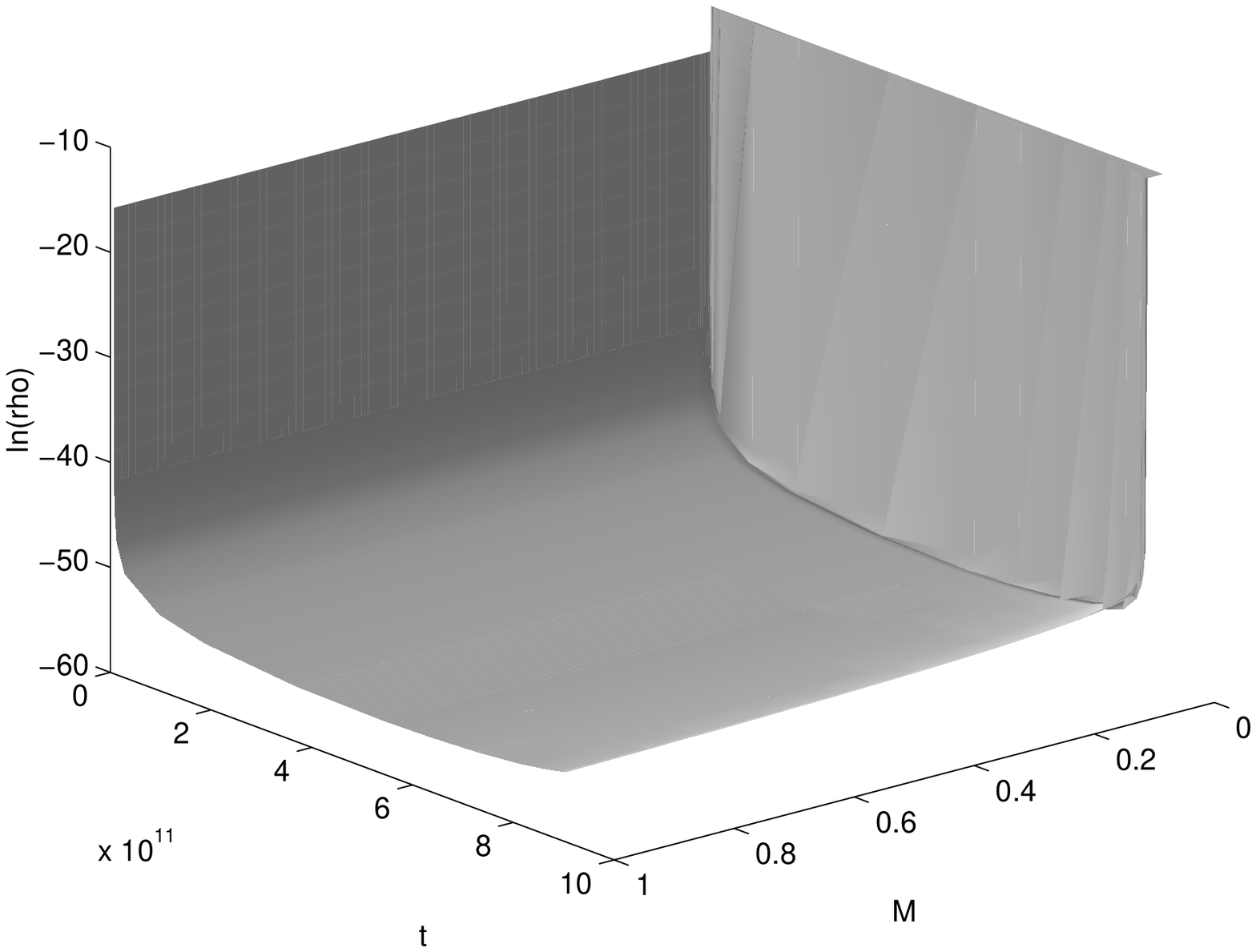}
 \caption{ \label{WHV0D30figB}
 The evolution of the
 L-T model for the formation of a galaxy around a
 pre-existing central black hole.  Shown are the evolution of $R(t,M)$ and
of $\rho(t,M)$ for run Vi0$\rho$f30.  In the $R(t,M)$ graph, on the left
there is no origin, rather a wormhole with a tiny lifetime
 re-collapses to zero size.  The thin flat wedge along the left is the
growing singularity.  (The apparent ripples are small numerical variations
overemphasised by the graphics program.)  In the $\rho(t,M)$ graph, again
with the view rotated by 180$^\circ$ relative to the $R(t,M)$ graph, the
density diverges towards the singularity, and the flat wedge represents
the part that has already collapsed.  By $t_2 =$ today, the black hole
mass has increased to $3 \times 10^9~M_\odot$.  Note that the graphs have
been clipped at $\log(R) = 0$ and $\log(\rho) = -5$ (geometric units).}
 \end{figure*}

   Our second model uses the identical exterior, but the interior is a
full
 Kruskal-Szekeres type black hole containing a temporary `wormhole', as
described by (\ref{WHEtB}), with parameters determined by the matching
(\ref{EtBWHmatch}). The same geometric units were used, and, using the `$-$'
sign in (\ref{EtBWHmatch}), the parameters are:
 \begin{multline}\label{MatchedWmHlParams}
   a = -4.7475 \times 10^{-8} ~,~~~~
   b = 0.012409 ~, \\
   t_{B0} = 0.0017385 ~,~~~~
   M_{\text{min}} = 2.0562 \times 10^{-11}
 \end{multline}
(Using the $+$ sign in (\ref{WHEtB}) gives $M_{\text{min}} = 0.06 > M_{BH} =
0.03$, which is not acceptable.) Figs \ref{WHV0D30figA} and \ref{WHV0D30figB}
show the arbitrary functions and the behaviour of the combined model for this
scenario.
 %
 %
 Because the exteriors are identical, the fluctuations of density and
velocity at recombination ($t_1$) outside $M_{BH}$ are again well within
CMB limits.  The wormhole mass (minimum in $M$) is $M_{\text{min}} =
2.0562~M_\odot$, and the future singularity first forms at $T_0 = 6.3613
\times 10^{-5}$~secs after the past singularity.  (The future and past
black hole singularities are the extension of the crunch and bang into the
middle of the wormhole.) The very short lifetime of the wormhole is a
consequence of the need for $E$ to go from $-1/2$ all the way up to $-
1.7669 \times 10^{-9}$ and arrive there with a negative gradient. (Even at
constant $M$, (\ref{ScaleT}) implies $T(M_{BH})/T_0 > 10^{13}$, and the
non-zero change in $M$ only increases this factor.) 
 Though models could no
doubt be found with quite different wormhole lifetimes, this example very
effectively highlights the fact that the nature of the central black hole
is essentially unknown.  By recombination ($t_1$), this black hole has
accreted%
 \footnote{
 The mass within the AH at $t_1$ is found by numerical root finding, using
(\ref{msmbhat2}) with $t_1$ instead of $t_2$ and (\ref{timeleaveAHcoll})
 }
 $246200~M_\odot$ within the apparent horizon, which is only
$0.0048586$~AU across.  Any 
 effect this might have on the CMB will not be 
observable for a long time.

 \section{Summary and conclusions}

 We have demonstrated the
 non-linear evolution of an initial density perturbation at recombination
into a galaxy with a central black hole at the present day, using the
spherically symmetric
 L-T model.  This is an application of the methods developed in papers I
\& II, in which an initial and a final state
 --- each a density profile or a velocity profile
 --- can be used to derive the arbitrary functions of an
 L-T model that evolves from one to the other.  To correctly describe this
process, a relativistic approach is necessary because Newtonian models are
inadequate for the description of black holes and their use inevitably
leads to conceptual inconsistencies and contradictions.  The
 L-T model is ideal for this purpose, as it has both Schwarzschild and
 Robertson-Walker limits, and a single model can describe a cosmology
containing a black hole.

   For the final state at $t_2 = 14 \times 10^9$ years $\approx$ today, we
chose the model of the mass distribution in the M87 galaxy used in
astronomical literature, eqs. (\ref{m87rpro}) and (\ref{arcvsdis}). More
exactly, we approximated this mass distribution by a more elementary
function whose values do not differ much in the range of interest, eq.
(\ref{myrprof}), so that $\rho$ can be calculated as an elementary
explicit function of the mass within a sphere of radius $r$.  M87 was
chosen since it is believed to contain a large black hole around its
center, and several of its parameters have been measured or calculated.
The inital fluctuation, at the recombination epoch ($t_1 = 10^5$ years),
was chosen to be consistent with limits from the CMB, even though the
smallest scales currently observable are much larger than those relevant
to galaxy formation.  We assumed zero initial velocity perturbation, i.e.
a Friedmannian velocity profile.  This was sufficient for a unique
numerical identification of an
 L-T model that evolves the given initial state into the given final
state.  The resulting
evolution of the
 L-T model, was found to be entirely reasonable
 --- the implied initial density amplitude was well within the
observationally allowed limit of $10^{-5}$, and the model was
 elliptic and already recollapsing by $t_2$ in  the whole range of
interest.  Assuming the presence of a central black hole today, these profiles
were taken to be valid for the exterior of the horizon, and a black hole
model was smoothly joined on as the interior.

   A theoretical model of a black hole must necessarily include the
accompanying entities: the final singularity, the apparent horizon and
(whenever appropriate) the event horizon. We first discussed the general
properties of the apparent horizon. It must necessarily exist in every
 L-T model. The future apparent horizon AH$^+$ exists in every collapsing
model, the past apparent horizon AH$^-$ exists in every expanding
 L-T model; the expanding and recollapsing model has both AHs that can
intersect each other only if there exists a neck or belly at which $R' =
M' = E' = t'_B = 0$, $E = -1/2$.  In every case, the 
 AH is given by the
equation $R = 2M$ that implicitly defines a function $t = t_{AH}(r)$. We
discussed whether the AH can be timelike, null or spacelike while going in
or out. It turned out that only two cases are excluded: outgoing timelike
for AH$^+$ and ingoing timelike for AH$^-$. The condition for the AH to be
nontimelike is $\ell \tdil {t_{AH}} r \leq \tdil {\cal M} r$, where $\ell
= \pm 1$ for AH$^\mp$ and ${\cal M}$ is the sum of rest masses of 
particles within the $r$-sphere. 
 For the Friedmann models, the AH is timelike
everywhere.

   The event horizon (EH) does not always exist; for example it is absent
in the $\Lambda = 0$ Friedmann models. When it exists, the EH$^+$ (EH$^-$)
is the collection of those future-directed outgoing (ingoing) radial null
geodesics that approach the AH$^+$ (AH$^-$) asymptotically as $t \to +
\infty$ ($t \to -\infty$). The equation of null geodesics is in general
intractable by exact methods, and so locating the EH inevitably requires
numerical integration. This can only be done case by case, for specific
forms of the L-T functions $E(r)$ and $t_B(r)$. With a space of infinite
volume, the numerical identification of the EH can most easily be done
after the spacetime is compactified so that the future (past) edge of the
AH$^+$ (AH$^-$) has finite time and radial coordinates.

   Then we calculated all the characteristic quantities of a black hole
(the past and future singularity, both AH$^-$s, radial null geodesics and
the EH+) in a simple illustrative toy model with $E < 0$ and displayed
them in spacetime diagrams. The model is recollapsing, has infinite total
mass and volume, and a duration between bang and crunch that is finite at
every finite mass, but goes to infinity as $M \to \infty$ and $E \to 0$.

   Since no observational data exist (and, presumably, will not exist for
a long time) concerning the interior of the horizon, two distinct forms of
this central black hole
 --- both
 L-T models
 --- were considered, firstly a condensation that collapses to a
singularity, and secondly a full
 Schwarzschild-Kruskal-Szekeres type wormhole topology.  The parameters of
these models were determined by matching their
 L-T functions to those of the exterior
 galaxy-forming model.  In the case of collapse to a black hole, the
central singularity is $\approx 4 \times 10^8$ years old by now.  For the
wormhole case, the final singularity forms almost immediately after the
Big Bang ($6.36 \times 10^{-5}$ sec.), and is by today about as old as the
matter in the galaxy. In this model, the black hole accretes mass very
fast into a very small volume, so that by recombination it had swallowed
up 246 380 $M_{\odot}$ in a region of diameter 0.00486~AU.  However, all
the numbers are strongly
 model-dependent, and there are no reliable observational constraints for
model selection.  The initial black hole is too small to have an
observable effect on the CMB.  Therefore we find both types of black hole
are possible.  Perhaps small black holes, that avoid evaporation by rapid
accretion, may seed galaxy formation.

   In fact, existing observational data does not have the resolution to
constrain the initial data for our model. For example, as shown in Paper
I, the perturbations of isotropy of the CMB temperature corresponding to
single galaxies should have the angular size of $\approx 4 \times 10^{-3}$
degrees, while the most precise current measurements have the resolution
of $0.2^{\circ}$.

   Then, in the vicinity of the apparent horizon, the geometry of
spacetime becomes measureably non-Minkowskian, while all the observational
data available on mass distribution within galaxies were calculated by
purely Euclidean reduction methods. We also stressed that what can be
inferred from observations is only the upper limit of the mass inside the
apparent horizon at a given time. It does not make sense to even speak of
an event horizon in the observational context.

   In view of the paucity of data, our approach was the first exploratory
step into an uncharted territory rather than an actual model to be
compared with observations.

   The main limitation of the spherically symmetric
 L-T model is the lack of rotation, which slows collapse and stablises
structures.  Thus the model is good for much of the evolution into the
 non-linear regime, but becomes less realistic as collapse sets in.

   Our results show that the
 L-T model is a very useful tool for this kind of
investigation. However, for its parameters to be
 fine-tuned to results of
observations, the observational data would have to be
 re-interpreted against the
background of the
 L-T geometries.

 \begin{acknowledgements}
 The research of AK was supported by the Polish Research Committee 
grant no 2 P03B 12 724.  CH thanks the South African National Research 
Foundation for a grant.
 \end{acknowledgements}

 \appendix

  \section{The Friedmann limit}

   We briefly specialize the above results to the Friedmann limit, where
 \begin{multline}\label{Frarbfuns}
   M = M_0 r^3, \qquad 2E = -kr^2, \\
   t_B = {\rm const}, \qquad R = rS(t),
 \end{multline}
 $M_0 > 0$ and $k > 0$ being arbitrary constants, and $S(t)$ being the
scale factor.  The apparent horizon, where $R = 2M$, has the equation
 \begin{equation}\label{AHink>0}
   S(t) = 2M_0 r^2.
 \end{equation}
 so by (\ref{DefOfB}), (\ref{dtdr_n}) and (\ref{timelikeAH}) we obtain
 \begin{gather}
   B = 1 - \frac{2M'}{R'} = -2 , \label{Frslope1} \\
   t'_n = j \frac{S}{\sqrt{1 - kr^2}\;} , \\
   t'_{\text{AH}^\pm} = \ell \frac {4 M_0 r^2}{\sqrt{1 - kr^2}\;}
      = 2 \ell j \, (t_n')_{\text{AH}^\pm} \, . \label{Frslope3}
 \end{gather}
 Hence, in the Friedmann limit both branches of the AH are entirely
timelike (outgoing in the expansion phase, incoming in the collapse phase) and
monotonic with $r$.  The derivatives $t'_{AH\pm}$ and $t'_n$ seem to become
infinite at $r = 1/\sqrt{k}$. This is a coordinate effect. As seen from the
metric ${\rm d}s^2 = {\rm d}t^2 - {S^2(t)} \left(\frac {{\rm d} r^2} {1 - kr^2}\
+ r^2 {\rm d} \omega^2\right)$, there is a coordinate singularity at $r =
1/\sqrt{k}$. Both derivatives become finite when the coordinates are changed so
that e.g. $r = \sin r'$. The quantity $B$ in eq. (\ref{Frslope1}) does not
depend on the choice of $r$.

For a completely general Robertson-Walker model (i.e not just $p = 0$),
repeating the whole reasoning, we obtain for the slope:
 \begin{equation}\label{slopAHRW}
  B = \frac{(\dot{S}^2 + k)}{S \ddot{S}},
 \end{equation}
 which, after making use of the $\Lambda = 0$ Einstein equations, is
equivalent to
 \begin{equation}\label{slopAHprho}
   B = \frac{- 2 \rho}{3 p + \rho}.
 \end{equation}
 This makes the AHs timelike for $1/3 > p/\rho > - 1/3$, but spacelike for
$p/\rho > 1/3$, or for $p/\rho < -1$.  (Note that $B = 0$ requires $\rho =
0$ or divergent $p$.)  Hence, it cannot be decided whether the AH is
timelike or not without knowing the precise shape of the function $S(t)$
(i.e. knowing the equation of state). It can only be said that as long as
the source in the Einstein equations is ordinary matter known from
laboratory (no cosmological constant or other
 self-accelerating medium), we will have $\ddot {S} < 0$. Consequently, by
(\ref{slopAHRW}), $B < 0$, which means that the AH will be outgoing in the
expansion phase and incoming in the collapse phase.

 \end{document}